\documentclass[12pt]{article}

\usepackage{graphicx}

\usepackage{color}
\usepackage{amsmath}
\usepackage{amsfonts}
\usepackage{amssymb}

\begin{document}

\title{Numerical evidence for relevance of disorder in a Poland-Scheraga 
DNA denaturation model with self-avoidance:\\ Scaling behavior of 
average quantities.}

\author{Barbara Coluzzi$^a$
\footnote{Corresponding author. Email: \tt coluzzi@lmd.ens.fr}
\hspace{.2cm} and Edouard Yeramian$^b$}

\maketitle

\begin{center}

a) {\em CERES-ERTI (Plateforme Environnement)},\\
 Ecole Normale Sup\'erieure, 24 rue Lhomond, 75005 Paris 

b){\em Unit\'e de Bio-Informatique
structurale},\\ Institut Pasteur, 25-28 rue du
Docteur Roux, 75724 Paris cedex 15\\

\begin{abstract}
\noindent
We study numerically the effect of sequence heterogeneity on the 
thermodynamic properties of a Poland-Scheraga model for DNA denaturation 
taking into account self-avoidance, {\em i.e.} with exponent $c_p=2.15$ 
for the loop length probability distribution. In complement to previous 
on-lattice Monte Carlo like studies, we consider here off-lattice numerical
calculations for large sequence lengths, relying on efficient algorithmic 
methods. We investigate finite size effects with the definition of an 
appropriate {\em intrinsic} length scale $x$, depending on the parameters of 
the model. Based on the occurrence of large enough {\em rare regions}, for a 
given sequence length $N$, this study provides a qualitative picture for the 
finite size behavior, suggesting that the effect of disorder could be sensed 
only with sequence lengths diverging exponentially with $x$. We further look
in detail at average quantities for the particular case $x=1.3$, ensuring 
through this parameter choice the correspondence between the off-lattice and 
the on-lattice studies. Taken together, the various results can be cast in a 
coherent picture with a crossover between a nearly {\em pure system like} 
behavior for small sizes $N \lesssim 1000$, as observed in the on-lattice 
simulations, and the apparent asymptotic behavior indicative of disorder 
relevance, with an (average) correlation length exponent 
$\nu_r \ge 2/d \:(=2)$.
\end{abstract}

\end{center}

\noindent
PACS: 64.60.Fr \hspace{.3cm} Equilibrium properties near critical points, 
critical exponents \\
PACS: 82.39.Pj \hspace{.3cm} Nucleic acids, DNA and RNA bases \\
PACS: 02.60.Cb \hspace{.2cm} Numerical simulation, solution of equations

\newpage

\section{Introduction}
\noindent
The discovery of the DNA double-helical structure, some 50 years ago, 
motivated the elaboration of the helix-coil model to account for the 
separation of the two strands, on physical bases \cite{review,PoSh1,RiGu}. 
The importance of this model from the biological point of view is obvious, 
since processing of the genetic information involves precisely the separation 
of the strands. Of course, under physiological conditions, the opening of the 
double-helix is not under the effect of temperature, but the differential 
stabilities in DNA sequences, as revealed by helix-coil analysis, could be 
sensed by biological effectors, such as proteins, under various types of 
constraints. The successful development of the helix-coil denaturation model 
required appropriate elaborations for the physics and the algorithmics, 
allowing accurate tests through comparisons with experimental data (melting 
curves). This field, very active in the sixties and seventies, has benefited 
recently from a renewed interest both from the biological side, for example in 
the context of genomic analysis, and from the physics side, notably in 
relation with questions relevant to the order of the transition in the 
homogeneous case and the effect of sequence heterogeneity. In the light of 
these still debated issues, both from the theoretical and the numerical 
points of view, the main focus of the present work is the numerical 
investigation of the relevance of disorder in a realistic DNA denaturation 
model {\em \`a~la} Poland-Scheraga, in which self-avoidance between loops and 
the rest of the chain is also taken into account. In what follows, before 
further detailing the particular system considered and the open questions, we 
first recall briefly the general background in terms of biological models, 
numerical methods and previous results.

\vspace{0.5cm}
\noindent
\textit{Basics for DNA denaturation:}
\noindent
DNA denaturation is an entropy driven transition, in which at some critical 
temperature $T_c$ the energy loss $\Delta E$ with the opening of base pairs 
is compensated by the entropic gain $ T \Delta S$ associated with the 
increased number of configurations accessible to the separated single 
strands. Experimentally, it is found that $T_c$ depends on different factors, 
in particular the $pH$ of the solution and the GC composition of the sequence, 
related to the ratio of the Guanine-Cytosine, GC, pairs to the 
Adenine-Thymine, AT, pairs. For homogeneous sequences, for $pH\sim 7$, 
typical values for $T_c$ are $T_{c,{GC}} \sim 110 \:^0 C$ and 
$\: T_{c,{AT}} \sim 70 \:^0 C$, respectively for GC and AT cases. Such 
differences reflect of course the fact that the pairing of Guanine to 
Cytosine involves three hydrogen bonds whereas that of Adenine to Thymine 
involves only two. 

For a given  biological sequence of length $N$, here identified, following AT 
and GC pairs, by the coupling energies $\{ \epsilon_i,  i=1,\dots,N \}$,
the denaturation transition can be followed with UV absorption. 
Correspondingly, the fraction $\theta_\epsilon(T,N)$ of closed base pairs, 
which is the order parameter of the transition in the thermodynamic limit 
$N \rightarrow \infty$, can be measured in such experiments based on 
differential absorptions for closed and open base pairs. The resulting curves 
display usually multi-stepped structures, with abrupt variations on small 
(sequence-depending) temperature ranges around $T_c$. Therefore, for a 
biological sequence of fixed length, the finite size order parameter 
$\theta_\epsilon(T,N)$ varies from zero to one (associated with complete 
denaturation), with a sequence-dependent behavior. Accordingly, the derivative 
with respect to temperature, $-d\theta_\epsilon(T,N)/dT$, displays typically a 
series of sharp peaks.

From the theoretical point of view, modeling DNA denaturation was 
essentially following two main directions: 1) for biological applications, 
in relation with melting experiments (sixties, seventies), sequence-dependent 
algorithmic elaborations for the handling of realistic physical models 
\cite{PoSh1,FiFr,Yeetal}, concerning notably the representation of 
denaturation loops, and, 2) for the study of the underlying physics, detailed 
characterizations of the properties for pure systems, neglecting 
sequence-specificity \cite{PoSh2,Fi1,CoMo,ThDaPe,KaMuPe1,HaMe,KaMuPe2,CaCoGr,CaOrSt,BaCaKaMuOrSt,GaMoOr}. 

\vspace{0.5cm}
\noindent
\textit{Physics of DNA denaturation for homogeneous sequences:}
\noindent
DNA denaturation is understandable in the framework of {\em almost 
unidimensional} systems \cite{Fi2}, and it is therefore associated with a 
peculiar kind of transition. In fact, the first models displayed no 
thermodynamic singularity \cite{PoSh1}, as they corresponded to $1d$ Ising 
models with only short-range (nearest-neighbor) interactions, with open and 
closed base pair states represented by an Ising spin.  
It was subsequently shown, notably by Poland and Scheraga \cite{PoSh2} (PS, in 
what follows), that the observed denaturation behavior can indeed be
described in terms of a simple $1d$ model, the helix-coil model,  
that consists of alternating regions of contiguous 
open base pairs (coiled regions or {\em loops}) and double-stranded 
ones (helical {\em segments}). In this model the transition in 
the thermodynamic limit is made possible through the adoption of 
appropriate long-range entropic weights for the single-stranded loops. 

More recently, several other models have been considered and studied, 
using in particular more realistic potential forms between base pairs 
\cite{CoMo,ThDaPe}. Since sharp transitions are observed experimentally,
with abrupt changes in $\theta_\epsilon(T,N)$ on small temperature ranges, 
it is expected that a model, accounting correctly for such results, should 
undergo a first order transition in the pure case. Indeed, this point has been 
studied rather extensively recently 
\cite{RiGu,CoMo,ThDaPe,KaMuPe1,HaMe,KaMuPe2,CaCoGr,CaOrSt,BaCaKaMuOrSt,GaMoOr,BuYo}.
In particular, it was demonstrated \cite{KaMuPe1} that the transition is 
of first order in pure PS models in which excluded volume effects for loops 
are not only with themselves, but also with the rest of the chain. Notably, 
with the probability distributions for loops with lengths $l$ at the critical 
point following a power law, $P(l,T_c) \propto 1/l^{c_p}$, the transition is 
of first order for $c_p$ exponents larger than 2 \cite{PoSh2,Fi1,Fi2}
(see also \cite{RiGu}). It was shown that in three dimensions, with the two 
strands described as self-avoiding walks (SAWs), the value for the exponent is 
$c_p\simeq 2.15$ \cite{KaMuPe1,KaMuPe2,CaOrSt,BaCaKaMuOrSt}. In comparison, 
$c_p=3/2$ for random walk (RW) loops \cite{PoSh2} and $c_p=1.76276(6)$ for 
SAW loops, with excluded volume interactions with the rest of the 
chain neglected \cite{Fi1,BeNi}.

\vspace{0.5cm}
\noindent
\textit{Biological and algorithmic backgrounds for sequence-specific DNA 
denaturations:}
\noindent 
The algorithmic problem was initially encountered for the implementation of 
sequence-specific calculations allowing notably experimental/theoretical 
comparisons in the study of melting curves. It seemed natural, in the 
beginning, to resort to transfer matrix formalisms as developed in physics 
because of the Ising-type formulation of the problem \cite{PoSh1}. Indeed, 
neglecting loop-entropy long-range effects, the calculation of the partition 
function for a sequence of size $N$ can be expressed simply as the product of
$N$ $2$x$2$ matrices. The extension to realistic models was at first handled 
through extended transfer matrices, of sizes growing up to $N$x$N$, for 
the proper description of interactions throughout the lengths of the 
sequences \cite{PoSh1}. Because of calculation burdens associated with 
such matrix sizes, alternative formulations were sought for the implementation 
of realistic models with affordable computation times. Representing the 
culmination of a series of developments, through some twenty years, an
appropriate algorithmic solution was proposed in 1977 by Fixman and Freire 
\cite{FiFr} (FF method), in which calculation efficiency was not at the 
price of oversimplifications in the physics, but relied instead on the 
numerical representation of the long-range effect as a multiexponential 
function. In this formulation, the time complexity for the evaluation of a 
complete denaturation map for a sequence of length $N$ is essentially 
proportional to $N$. This reduced complexity is to be compared with the 
intrinsic complexity of the model scaling as $N^3$, if we were to consider 
exact one-way calculations along the sequence. 

In this background, no generalizations were proposed for the ideas in the FF 
for a long period of time, possibly because of the formulation of this method 
in the prolongation of an algorithm by Poland \cite{PolAlgo}, expressed in the 
rather specialized context of conditional probabilities recursions specific to 
the linear DNA helix-coil model. As a matter of fact, the only applications of 
the FF concerned the implementation of the algorithms in computer 
programs, for DNA melting calculations (such as in the POLAND \cite{PolProgr}
or in the MELTSIM programs \cite{Bl}). However, upon revisiting the original 
derivations, it appears that the idea associated with the 
multiexponential representation, relying on the fundamental property of 
exponential function, corresponds to a powerful concept amenable to many 
generalizations for realistic models with long-range effects. Accordingly, 
based on explicit partition function calculations, the SIMEX (SIMulations with 
EXponentials) method was first derived as a reformulation of the FF for the 
linear helix-coil model \cite{Yeetal}, with further generalizations to 
higher-order models involving several, mutually coupled, long-range effects 
\cite{Ye1,GaOr}. For such systems, with two or more long-ranges, the 
reductions of complexities by several orders of magnitudes can be associated 
with calculation times reduced by million folds. The basic concepts for 
higher-order models were originally illustrated with a circular 
DNA helix-coil model-problem involving two long-range contributions 
\cite{Ye1}. The corresponding principles were further transposed to a 
linear helix-coil model with non-symmetrical loops \cite{GaOr}.

On the experimental side, linear helix-coil models were successfully compared 
with experimental denaturation curves \cite{Bl}. In the beginning, one of the 
motivations in the elaboration of DNA denaturation models was to ask the 
question of possible relations between genetics (coding/non-coding) and 
physical (helix/coil) segmentations, because of the importance of the 
separation of the two strands in the processing of the genetic information. 
With little genomic sequences available at the end of the seventies and 
beginning of eighties, no clearcut conclusions were reached for such 
relations. More recently, with the availability of complete genomes, it was 
possible to resume such investigations on much larger bases, demonstrating 
variable correspondences between the two types of segmentations, depending on 
the genomes \cite{Ye,YeBoLa,YeJo,CaMaBl}. For genomes with very sharp 
correlations it was even possible to propose {\em ab initio} gene 
identification methods purely on physical bases \cite{YeBoLa}.

\vspace{0.5cm}
\noindent
\textit{Physics of DNA denaturation with disorder:}
\noindent
On the physical side, DNA denaturation models are associated with important 
open questions such as, notably, the relevance of sequence-heterogeneity for 
the thermodynamic limit behavior. At the beginning of the seventies, it was 
noted by Poland and Scheraga \cite{PoSh1} that ``sequence heterogeneity 
dramatically broadens the transition'', but it is only recently that such 
problem has been addressed on more rigorous bases and the transition in 
disordered PS models with $c_p>2$ was also investigated 
\cite{GaMo,MoGa,Co,GiTo1,GiTo2,To}. Indeed, in the homogeneous case, such 
systems exhibit peculiar first order transitions characterized by a diverging 
correlation length, and it is therefore not clear to which extent general 
theoretical results on the effect of disorder \cite{Ha,AiWe,Be} can be 
applied. The question has been addressed both with analytical 
\cite{GiTo1,GiTo2,To} and numerical approaches, with either off-lattice 
\cite{GaMo,MoGa} or on-lattice \cite{Co} implementations in this latter case. 
It appears however hard to reconcile these various results. The off-lattice 
studies of \cite{GaMo,MoGa}, involving very large chain lengths, suggest a 
peculiar transition, of first order as in the pure case but not obeying usual 
finite size scaling and exhibiting two different correlation length exponents, 
associated respectively with {\em typical} and {\em average} quantities. On 
the other hand, the Monte Carlo like numerical simulations in \cite{Co}, 
limited to small chain lengths, agree with a second order transition in the 
presence of disorder, though it was not possible to rule out completely a 
transition still of first order. Finally, from the analytical standpoint, 
under quite general hypotheses encompassing PS models with $c_p>2$, it was 
shown that the transition is expected to be at least of second order and 
possibly smoother \cite{GiTo1,GiTo2,To}. 

\vspace{0.5cm}
In the background above, in addition to their relevance to experimental DNA 
denaturation, PS models with sequence heterogeneity represent interesting 
toy-models for addressing general open questions relative to the proprieties 
of random fixed points. The detailed study of such systems could also help 
elaborating the correct approaches to be used in the interpretation of data on 
disordered models. In this direction, we perform here a numerical analysis of 
a disordered PS model with $c_p=2.15$. Relying on appropriate algorithmic 
formulations (SIMEX) we consider long sequences. With the definition adopted 
for the model, and the choices for the parameter values, the calculations are 
made directly comparable with previous on-lattice results \cite{Co}. In 
addition, the observed behavior should be also related to that found in 
the previous off-lattice studies of a different disordered PS model with 
$c_p=2.15$ \cite{GaMo,MoGa}, in which the same multiexponential representation 
for the long-range loop entropy law was adopted.

Our findings show the existence of very strong corrections to scaling. 
Moreover it appears that, for a given size, the effect of disorder is 
qualitatively described by an appropriately defined {\em intrinsic} length 
scale $x$ depending on model parameters. These observations provide a possible 
explanation for the discrepancies between previous results 
\cite{GaMo,MoGa,Co}, as well as for an apparent dependence of the evaluated 
critical exponents on model parameters noted in \cite{Co}. In fact, in the 
frame of the picture proposed here, the size at which the effect of disorder 
becomes evident could diverge exponentially with $x$. More precisely,
for the value $x=1.3$ chosen for the present detailed study, it is
possible to observe a crossover between a nearly {\em pure system like}
behavior, consistent with the one observed in simulations \cite{Co}, and the 
apparent asymptotic one. With corrections to scaling taken into account, the 
model clearly displays a smooth transition, corresponding to a value for the 
correlation length exponent $\nu_r \ge 2/d \:(=2)$. Nevertheless, since our 
results refer to average quantities, they do not rule out the possibility 
suggested in \cite{GaMo,MoGa,Mopr,MoGapr} of a transition governed by two 
different correlation lengths. The analysis for the clarification of this 
point is left for a forthcoming work \cite{CoYe2}.
 
\section{Models {\em \`a la} Poland-Scheraga} 
\subsection{The pure case and the role of the exponent $c_p$}
\noindent
Pure PS models for DNA denaturation are described in a rather extensive 
literature, and in particular the ingredients which make the 
transition of first order are discussed in several recent works 
\cite{RiGu,KaMuPe1,HaMe,KaMuPe2,CaCoGr,CaOrSt,BaCaKaMuOrSt,GaMoOr}.
Here we will only recall results for linear PS models 
with symmetric loops, in which one fully takes into account self-avoidance
through an appropriate choice of the loop length distribution probability
exponent $c_p$. 

The position of a base pair along the sequence is labeled by $i$ 
($i=1,\dots,N$) and its configurational state is represented by $s_i$,
with $s_i=1$ for a closed pair and $s_i=0$ for an open pair. 
In the corresponding on-lattice representation, the two strands in the model 
can be visualized as two interacting RWs with the same origin in $d$ 
dimensions. A pair is in the closed state if and only if the two bases 
are in the same position $i$ along the two strands and occupy 
the same lattice point \cite{CaCoGr} (RW-DNA). One can write the canonical 
partition function for the system in the form:
\begin{equation}
Z_N=\sum_{E} {\cal N}(E) e^{-\beta \: E},
\label{z1}
\end{equation}
where ${\cal N}(E)$ is the number of configurations with the same total 
energy $E$ and $\beta=1/T$ is the inverse temperature, taking Boltzmann 
constant $k_B=1$ for simplicity.

The contribution of a single base pair in the closed state to the total 
energy $E$ is $\Delta E=-\epsilon$, independent from the position $i$ in the 
pure case. The number of configurations of a closed segment of length
$n$ increases as $\mu^{n}$ and therefore its contribution to the total entropy 
$S(E)=\log {\cal N}(E)$ is $\Delta S=n \log \mu$. Here $\mu$ is a parameter of 
the model, interpretable as the on-lattice connectivity constant ($\mu=2d$ for 
$d$-dimensional RWs on a cubic lattice). A denatured region of length $l$ 
is associated with a single-stranded loop of length $2l$, and the 
corresponding number of configurations is given by $\mu^{2l}/(2l)^{c_p}$. The 
factor $1/(2l)^{c_p}$ takes into account the fact that the two separated chains 
have to meet again at some distant point, and the relation $c_p=d/2$ 
holds for $d$-dimensional RWs. 

Assuming that the first and the last base pairs are always coupled, a given 
configuration of the system is described by the lengths of the closed segments 
$\{ n_j \}$ and by those of the denatured loops $\{ l_k \}$, 
with $\sum_j n_j + \sum_k (l_k-1) =n_{tot}+l_{tot}=N$. Its energy is 
$E=- \epsilon \sum_j n_j=-\epsilon n_{tot}$, only depending on $n_{tot}$.  
Therefore, in the partition function, the degeneracy factor ${\cal N}(E)$ 
includes both the entropic contribution of segments $\mu^{n_{tot}}$ and that
of loops  $\mu^{2l_{tot}} \prod_k {1}/{(2l_k)^{c_p}}$, the latter being to be
summed over all the possible sets $\{ l_k>2 \}$ associated with a given 
total length $l_{tot}=\sum_k (l_k-1)$. Correspondingly, the partition function
can be also written as:  
\begin{eqnarray}
Z_N&=&e^{2N \log \mu} \sum_{n_{tot}} e^{n_{tot} (\beta \epsilon-\log \mu)} 
\sum_{\{l_k >2\}}  \delta(n_{tot}+l_{tot}-N)
\prod_k {\frac{1}{(2l_k)^{c_p}}}=\nonumber\\
&=&e^{2N \log \mu}
\sum_{\{ n_j \}} \sum_{\{l_k>2\}} \delta(n_{tot}+l_{tot}-N) 
\prod_j e^{n_j(\beta \epsilon-\log \mu)} \prod_k {\frac{1}{(2l_k)^{c_p}}} 
\label{zn}
\end{eqnarray}
It can be noted that the factor $e^{2N \log \mu}$ contributes an additive 
constant to the entropy and it is therefore not relevant for the description 
of the thermodynamics of the system. In what follows, we study properties of 
$ Z^*_N= Z_N/e^{2N \log \mu}$. Accordingly, the number of loops of length 
$l_k$ is taken equal to $1/(2l_k)^{c_p}$ and a negative contribution to the 
entropy $\Delta S= -\log \mu$ is associated to each base pair in the closed 
state. It is seen qualitatively that the possible change in the thermodynamic 
limit behavior occurs at the temperature for which 
$\beta_c \epsilon \sim \log \mu$. From the last expression in (\ref{zn}) it is 
moreover clear that, in the computation of the grand canonical partition 
function ${\cal Z}$, the contributions of helical segments and loops are 
decoupled and one obtains a geometric series 
\cite{PoSh1,PoSh2,Fi1,KaMuPe2,Fi2}:
\begin{equation}
{\cal Z} = \sum_{N} z^N Z_N = 
\sum_{\rho} \left \{ \left [ \sum_{n=1}^{\infty} z^{n} e^{n 
(\beta\epsilon - \log \mu)} \right ] \left [ \sum_{l=2}^{\infty} 
{\frac{z^l}{(2l)^{c_p}}}\right] \right \}^\rho =\frac{{\cal Z}_{S}{\cal Z}_{L}}
{1-{\cal Z}_{S} {\cal Z}_{L}},
\end{equation}
where we introduced the fugacity $z$ and 
${\cal Z}_S$ and ${\cal Z}_L$ refer to the {\em segment} (helical) and 
{\em loop} (coil) grand partition functions respectively:
\begin{eqnarray}
{\cal Z}_{S}& = &\sum_{n=1}^{\infty} z^n e^{n
(\beta\epsilon - \log \mu)} = \frac{ze^{(\beta\epsilon - \log \mu)}}{1-ze^{  
(\beta\epsilon - \log \mu)}},\\
{\cal Z}_{L}&=&\sum_{l=2}^{\infty} \frac{z^l}{(2l)^{c_p}}.
\end{eqnarray}

Since the behavior for $N \rightarrow \infty$ is dictated by the 
fugacity value $z^*$ corresponding to the pole nearest to the origin, the 
system undergoes a phase transition when a critical temperature $T_c$ is found 
below which the zero of the denominator becomes smaller than the smallest pole 
of the numerator. The possibility of the transition and its order both depend 
on the value of the exponent $c_p$ \cite{PoSh1,PoSh2,Fi1,KaMuPe2,Fi2}.
In detail, for $c_p < 1$ there is no thermodynamic singularity, whereas for 
$c_p > 1$ the following situations must be distinguished: a smooth transition 
for $1 < c_p < 3/2$ with a specific heat exponent $\alpha_p<0$, a second order 
transition for $3/2 \le c_p \le 2$ and finally a first order transition for 
$c_p > 2$. In fact, these distinctions can be understood considering the 
properties of ${\cal Z}_L$, {\em i.e.} those of the distribution probability 
of the loop length at the (possible) critical point $P(l,T_c)=1/(2l)^{c_p}$:
\begin{eqnarray}
\sum_l P(l,T_c) &=& \infty \mbox{ for } c_p \le 1 \\
\sum_l l P(l,T_c) &=& \left \{
\begin{array}{lcl}
\infty & \mbox{ for }& c_p \le 2. \\
const & \mbox{ for }& c_p>2
\end{array}
\right.
\end{eqnarray}
In the case $c_p \le 2$, the mean length $ \langle l \rangle$ for loops
at the critical point diverges and the system exhibits large coiled 
regions, in which most of the bases are involved. On the contrary, for 
$c_p>2$, the mean loop length at $T_c$ is finite and correspondingly it is 
possible to show that the density of closed base pairs 
$\theta(T,N)= \langle n \rangle /N$, and therefore the energy density,
varies abruptly in the thermodynamic limit, from the value zero at high 
temperatures to a finite value at $T_c$. In detail, when approaching the 
critical point from the low temperature (helical) phase, one finds 
\cite{PoSh1,PoSh2,Fi1,KaMuPe2,Fi2}:
\begin{equation}
\theta(T) \equiv \lim_{N \rightarrow \infty} 
\frac{1}{N} \sum_{i=1}^N \langle s_i \rangle \propto
\left \{
\begin{array}{lcl}
(T_c-T)^{\frac{2-c_p}{c_p-1}} & \mbox{ for } & 1 < c_p \le 2 \\
(T_c -T)^0  & \mbox{ for } & c_p > 2 \\
\end{array}
\right.
\label{orpa}
\end{equation}
For example, the RW-DNA model in three dimensions exhibits a second order 
transition with $\alpha_p=0$, since $c_p=3/2$, whereas in five dimensions 
it undergoes a first order transition, since $c_p=5/2$, as
confirmed both by exact computations of thermodynamic quantities and by 
on-lattice numerical simulations \cite{CaCoGr}. 

\subsection{Exponent value $c_p=2.15$}
\noindent
All interactions between different loops and helical segments are neglected in 
classical calculations of the grand canonical partition functions for 
helix-coil models. It is possible to account for self-avoidance of each loop 
with itself through the appropriate choice of the exponent 
$c_p=1.76276(6)$ \cite{Fi1,BeNi}, corresponding roughly to the value 
adopted usually for comparisons with experimental data \cite{Bl}. 
More recently, it was demonstrated that self-avoidance of the loops with the 
rest of the chain can be also taken into account, and that intriguingly the 
pure PS models exhibit first order transitions in this case 
\cite{KaMuPe1,HaMe,KaMuPe2}. The exponent $c_p$, corresponding to a 
self-avoiding loop embedded in a self-avoiding chain, 
can be predicted from conformal theory results \cite{Du}, and in particular 
it was found that $c_p\simeq 2.15$ in three dimensions, the
transition being of first order also in $d=2$. It is notable that such 
determination provides the appropriate value of $c_p$ to be used as an 
{\em input} in off-lattice calculations.

In the Monte Carlo like simulations, one studies an on-lattice model (SAW-DNA) 
in which self-avoidance is completely taken into account, by considering
two interacting SAWs, with two monomers allowed to occupy the same 
lattice point if and only if their positions along the two chains are 
identical, thus representing complementary base pairs \cite{CaCoGr}. 
In the pure $3d$ case, it was found that this system exhibits a first order 
transition, with the maximum of the specific heat diverging linearly with the 
chain length. It was subsequently shown \cite{CaOrSt,BaCaKaMuOrSt} that the 
value of the exponent describing the probability distribution for the loop 
lengths at the critical point is in perfect agreement with the theoretical 
prediction, $c_p \simeq 2.15$. An off-lattice pure PS model with $c_p=2.15$ 
was also studied numerically \cite{Sc}, finding the same scaling behavior than 
in $3d$ SAW-DNA, apart from the strong finite size corrections which appear to 
be more important in the on-lattice situation.

Even though of first order, the transition is characterized by a diverging 
correlation length, which can be identified from the behavior of $P(l,T)$ 
\cite{KaMuPe1,KaMuPe2,BaCaKaMuOrSt}:
\begin{equation}
P(l,T) \propto \frac{e^{-l/\xi(T)}}{(2l)^{c_p}}
\label{plt}
\end{equation}
with
\begin{equation}
\xi(T) \propto (T-T_c)^{-\nu_p} \mbox{ for } T \rightarrow T_c^-.
\label{corr}
\end{equation}
where $\nu_p=1$ for $c_p>2$ and $\nu_p=1/(c_p-1)$ for second order (or 
smoother) transitions. It can also be predicted that the free energy density 
$f(T)$ takes the value zero in the high temperature phase and that it 
behaves proportionally to $1/\xi(T)$ for $T \rightarrow T_c^-$, leading
again to the behavior of the energy density and of the order parameter for 
different $c_p$ values given in (\ref{orpa}). Therefore, the hyperscaling 
relation $\alpha_p=2-\nu_p$ is clearly fulfilled, both for $c_p \le 2$ and for 
the first order ($\alpha_p=1$) case. 

\subsection{Effect of disorder: previous results}
\noindent
Disorder is introduced to account for sequence-heterogeneity, with parameter 
values depending on the chemical nature of base pairs (AT or GC) at a given 
position along a sequence. There are a few studies on the effect of disorder 
on general properties of DNA denaturation models in which self-avoidance is 
neglected \cite{CuHwa,LuNe,ArSa,TaCh,KaMu}. Previous numerical works on 
disordered models {\em \`a la} PS was mainly for comparison of the 
predictions with experimental data and genetic signals 
\cite{Bl,Ye,YeBoLa,YeJo,CaMaBl} and for the study of the effect of base pair 
mismatches \cite{GaOr}, where one usually takes also into account the stacking 
contributions, with the coupling energies depending on the chemical nature of 
base pairs at positions $i$ and $i+1$. For comparisons with experimental 
melting curves, it is moreover necessary to take into account the possibility 
for complete dissociation of the two strands in the molecule 
\cite{PoSh1,IvZeZo}. We also notice that biological sequences exhibit 
long-range correlations and strong variability in GC compositions, both 
according to genomes and within chromosomes. Letting aside for the 
present such sophistication, we sum up some recent results for simple 
disordered models {\em \`a la} PS with self-avoidance, such as those 
considered in off-lattice \cite{GaMo,MoGa} and on-lattice \cite{Co} studies, 
which allow nevertheless to capture essential features of the effect of 
disorder.

In these various works, disorder enters only through the position 
dependent contribution of a closed base pair to the total energy. In detail, 
the $\{ \epsilon_i \}$ are quenched random variables distributed following a 
binomial probability, corresponding to GC composition equal to 1/2:
\begin{equation}
P(\epsilon)={\frac{1}{2}} \left [ \delta(\epsilon-\epsilon_{AT})
+ \delta(\epsilon-\epsilon_{GC}) \right ],
\label{peps}
\end{equation}
with $\epsilon_{AT} < \epsilon_{GC}$. One is interested in the 
thermodynamic properties of the quenched free energy density:
\begin{equation}
f(T)=\lim_{N \rightarrow \infty} f_\epsilon(T,N)=
\lim_{ N \rightarrow \infty}
\overline {f_\epsilon(T,N)} = - \lim_{N \rightarrow \infty}
{\frac{1}{\beta N}} \overline{ \log Z_{N,\epsilon}},
\end{equation}
where, as usual, $\overline{(\cdot)}$ denotes the average over disorder and 
$f_\epsilon(T,N)$ is a self-averaging quantity in these models \cite{GiTo1}.

Generally speaking it is known, from the theoretical point of view, 
that disorder can modify very significantly the fixed point of a system, and 
therefore its critical exponents. In what follows, the notations with 
subscripts $p$ and $r$ refer to the (possibly different) {\em pure} and 
{\em random} system fixed points, respectively. A series of results 
\cite{Ha,AiWe,Be}, and notably the well known Harris criterion \cite{Ha}, 
demonstrate that disorder is relevant as soon as the specific heat exponent 
fulfills the condition $\alpha_{p} > 0$. Correspondingly, in the presence of 
disorder, the transition becomes smoother and it is in particular expected that
$\alpha_{r} \le 0$, {\em i.e.}, from the hyperscaling relation, a 
correlation length exponent $\nu_r \ge 2/d$ \cite{ChChFiSp,AhHaWi,PaScZi}. 
It is however important to stress that these results are obtained 
essentially for magnetic systems and it is not clear to which extent they 
are relevant to the case here considered, concerning interacting polymers 
undergoing a first order transition characterized by a diverging correlation 
length. 

An analysis in terms of pseudo-critical temperatures \cite{WiDo}
was applied recently to disordered PS models with different $c_p$ values 
\cite{GaMo,MoGa,Mopr,MoGapr,GaMo2,MoGa2}. In these studies, an appropriate 
sample-dependent $T_c(\epsilon,N)$ was defined and measured, looking
in particular at the associated probability distribution. The results point 
towards irrelevance of disorder for $c_p=3/2$, corresponding to the marginal 
case $\alpha_p=0$. On the contrary, relevance of disorder was found for 
$c_p>3/2$. Importantly, peculiar behavior was observed for the value 
$c_p=2.15$, whereas the situation appears to be clear for $c_p=1.75$. 
Indeed, for $c_p=1.75$, compatible estimates for $\nu_r \sim 2.7 \: (>2/d=2)$ 
were obtained from the scaling of the average pseudo-critical temperature 
$T_c(N)$ and from that of (the square root of) its fluctuations 
$\delta T_c(N)$. In the case $c_p=2.15$, instead, a scaling of the average 
critical temperature $T_c(N) \sim 1/N$ was reported, suggesting a still first 
order transition with exponent $\nu_{r,1}=1$. However it was also found a 
scaling for the fluctuations following $\delta T_c(N)\sim 1/N^{1/2}$, which 
was associated to a different exponent $\nu_{r,2}=2 \: (=2/d)$. It was then 
suggested \cite{GaMo,MoGa,Mopr,MoGapr} that these observations are compatible 
with a system still exhibiting a first order transition but in which scaling 
laws are no more fulfilled, characterized by two different correlation 
lengths: a {\em typical} one, $\xi_1 \sim 1/|T-T_c|$, describing the 
behavior of a typical sample in the thermodynamic limit, with $\nu_{r,1}=1$, 
and an {\em average} one, $\xi_2 \sim 1/(T-T_c)^2$, describing the behavior 
of average quantities, dominated by rare fluctuations, with $\nu_{r,2}=2/d=2$. 
The resulting two-sided scenario is therefore that disorder is irrelevant to 
the typical sample and, in the same time, the obtainment of 
the $\nu_{r,2}$ value is in agreement with the theoretical expectations 
\cite{ChChFiSp,AhHaWi,PaScZi}. 

By contrast, usual finite size scaling analysis of Monte Carlo like 
simulations results on a $3d$ disordered SAW-DNA model (DSAW-DNA) \cite{Co}, 
suggested a transition governed by an (average) correlation length exponent 
$\nu_r \simeq 1.2$. However, in this study, average energy curves were 
observed to cross at the same point within the errors in the estimations, and 
accordingly the possibility $\nu_r=1=\nu_p$ could not be completely ruled out. 
The findings were further confirmed by the analysis of the behavior of 
$\overline{P_\epsilon(l,T)}$ at the critical point, which led to the 
compatible value $c_r \simeq 1.9 \simeq 1+1/\nu_r$ when considering the 
largest sizes and taking into account the presence of a finite correlation 
length. But again, particularly with the smallest chain lengths, estimations 
$c_r>2$, still compatible with a first order transition, were obtained. 
It is noticeable that the affordable sizes in Monte Carlo like studies are 
significantly smaller (factors of order $\sim 2000$) than the sequence 
lengths accessible to off-lattice recursive canonical partition 
function calculations for PS models.

Finally, in recent theoretical works based on a probabilistic approach 
\cite{GiTo1,GiTo2,To}, it was shown for a general class of interacting polymer 
models that the transition becomes at least of second order in the presence of 
disorder. The frame of this approach covers the PS models with $c_p>3/2$, 
including the case $c_p>2$ with corresponding first order transition in the 
pure system. Accordingly, these conclusions are expected to also cover the 
$3d$ DSAW-DNA case. Following such studies, it is expected that $\nu_r \ge 2$
both for average and typical quantities, though the possibility of different 
correlation lengths is not ruled out \cite{To}. On the other hand, according 
to other theoretical results in which self avoidance is neglected 
\cite{TaCh,KaMu}, disordered models {\em \`a la} PS could undergo a definitely 
smoother transition, corresponding to an essential singularity in the 
free energy. 

\subsection{A possible scenario for the finite size behavior} 
\noindent
Before presenting our numerical findings, it is worth discussing qualitative 
features expected, at fixed chain lengths $N$, for the behavior of disordered 
PS models. As previously recalled, disorder should be relevant to 
these models as soon as $c_p>3/2$. Moreover, on general grounds, one can argue 
that the behavior of a system near the transition point is governed by given 
critical exponents which do not depend on model details. Correspondingly, one
would expect that both the form of $P(\epsilon_i)$ and the precise choices for 
the parameters (here $R=\epsilon_{GC}/\epsilon_{AT}$, $\mu$ and GC 
composition) should not correspond to different thermodynamic limit 
singularities. On the other hand, such choices could have strong influence on 
finite size effects.

The disordered PS model in \cite{GaMo,MoGa} and the $3d$ DSAW-DNA in \cite{Co} 
involve the same $c_p=2.15$, either as a direct input for the recursive 
calculations or as consequence of the implementation of self-avoidance in the 
simulated model. Nevertheless, there is a first noticeable difference 
between the two systems studied\footnote{We thank Thomas Garel for pointing 
this difference to our attention}, as the off-lattice calculations were 
inspired from a wetting transition model \cite{MoGa2}, in which it is 
forbidden for two consecutive elements to be in the closed state 
simultaneously ({\em i.e.}, in our notation, if $s_i=1$ then $s_{i+1}=0$). 
Also the connectivity constant $\mu$ is not the same, since it was fixed to 
the value $\mu=2$ in \cite{GaMo,MoGa}, whereas it is an output of the model in 
the on-lattice simulations, and one finds $\mu \simeq 4.7$ for SAWs on a $3d$ 
cubic lattice. In addition, the two studies involved significantly different 
$R$ values. In \cite{GaMo,MoGa} the choice $R \simeq 1.098$ was adopted, for 
obtaining a critical temperature ratio $T_{c,GC}/T_{c,AT}$ close to the 
experimental value. On the other hand, in \cite{Co}, the value $R=2$ was 
studied in detail and the values $R=4$ and $R=\infty$ (corresponding to the 
choice $\epsilon_{GC}=1$ and $\epsilon_{AT}=0$) were also considered. In the 
latter study, preliminary results suggested that the (average) 
correlation length exponent $\nu_r$ could increase with $R$, ranging from 
$\nu_r\simeq 1.18$ for $R=2$ to $\nu_r \simeq 1.33$ for $R=\infty$. 

For proper understanding of these findings, and for a qualitative analysis of 
the expected finite size behavior, it can be important to consider in some 
detail the potential key role of {\em rare regions} in the generated 
sequences. Indeed, the possible presence of such regions, of large enough 
size, can explain the presence of strong corrections to scaling, which could 
therefore depend both on model details and on the precise choice for the 
parameters. It can be noted that temperature and disorder appear only in the 
$\pi_i=\exp[{s_i(\beta \epsilon_i-\log \mu})]$ terms in the partition 
function, which are clearly invariant under the transformation
$(\epsilon_i \rightarrow  \alpha \epsilon_i,\: T \rightarrow  \alpha T)$.
Moreover, in the pure system, the transition occurs around the temperature 
$T_{c,{p}} \sim \epsilon /\log \mu$, at which the energetic contribution for 
the two bound chains is of the same order than the entropic loss. In the 
presence of disorder, for a given sequence, it is expected to observe the 
multi-step behavior in $\theta_\epsilon(T,N)$ displayed by experimental DNA 
denaturation curves. This results from the presence of regions with 
different local contents in terms of GC to AT ratios, associated accordingly 
with different local melting temperatures. 

In the simplest extreme case, one imagines two regions $A$ and $B$, of about 
the same length $L$, completely dominated by AT and GC compositions 
respectively. In such situation, the local transition in region $A$ is driven 
by $\epsilon_{AT}$ energies, with local critical temperature 
$T_{c,loc}(A) \sim T_{c,{AT}} \sim \epsilon_{AT}/\log \mu$,
whereas the local transition in region $B$ is associated with the higher local 
critical temperature 
$T_{c,loc}(B) \sim T_{c,{GC}} \sim \epsilon_{GC}/\log \mu \sim R T_{c,loc}(A)$.
In this illustrative example, for a given temperature, the contributions 
in the partition function of total $\pi^{tot}_{A,B}$ factors 
corresponding to the configurations with base pairs in 
the closed state, will be significantly different for $A$ and $B$ regions. 
One obtains, for $T=T_{c,loc}(B)$ and $s_i=1 \:\: \forall i \in A,B$:
\begin{eqnarray}
\pi^{tot}_{A} & = & \prod_{i \in A} \pi_i =
\exp \left [{L (\beta_{c,loc}(B) \epsilon_{AT}-\log \mu)}
\right ] \sim O(1),
\nonumber \\
\pi^{tot}_{B} & = & \prod_{i \in B} \pi_i =
 \exp \left [ {L (\beta_{c,loc}(B) \epsilon_{GC}-\log \mu)}
\right ] \sim \exp(-L/x).
\end{eqnarray}
Since, whereas $\beta_{c,loc}(B) \epsilon_{GC} -\log \mu \sim 0$, one has
\begin{equation}
\beta_{c,loc}(B) \epsilon_{AT}-\log \mu \sim - \log \mu \frac{R-1}{R}
\equiv -\frac{1}{x},
\label{icsdef}
\end{equation}
defining the parameter $x$. From these expressions it is possible to argue 
that the larger the value of $L$ with respect to $x$ the more the effect of 
disorder will be {\em felt} by the finite size system, {\em i.e.} the 
difference between the weights of configurations corresponding to closed $A$ 
and $B$ regions in the partition function will be higher. On the other hand, 
the probability for such an extreme case in a particular sequence of size $N$ 
is quite small. With the choice for $P(\epsilon)$ in (\ref{peps}), large $N$ 
and $L$ values and $2^L \gg N \gg L$, the probability of $L$ contiguous 
elements of the same type is simply $\sim N/2^L$. Therefore this probability, 
though approaching 1 in the thermodynamic limit for any finite $L$, becomes 
rapidly negligible with increasing $L$ for fixed chain length $N$. Following 
these considerations, for the finite size system to feel the effect of 
disorder, the chain lengths necessary for observing {\em rare regions} with 
$L \gg x$ could be not reachable for large $x$ values. For more quantitative 
analysis, at least in the extreme case considered, let us suppose that at the 
length scale $N_1$ a region of size $L_1$ is observed for the parameter value 
$x_1$, such that $L_1/x_1 \gg 1$, with non negligible probability 
$N_1/2^{L_1}$. Then, in order to make the same observation for $x_2 > x_1$, it 
will be necessary to consider length scales of order 
$N_2 \sim N_1 \exp \left [ (x_2/x_1 -1) L_1 \log 2 \right ]$, thus involving
exponential increases with the ratio $x_2/x_1$ for the sequence lengths $N$.

For attempting to understand the role of the long-range loop entropic effect, 
one can impose that, at the temperature $T_{c,loc}(B)$, the weight of the 
configuration associated with region $A$ in the closed state is significantly 
smaller than that of the configuration associated with an open region
corresponding to a single loop (of size $L$), getting the condition 
$L/x \gg c_p \log L$.
Correspondingly, one can argue that, with increasingly larger $c_p$ values, 
increasingly larger {\em rare region} lengths will be necessary for observing 
cooperative melting behavior at different temperatures. Moreover, the 
considered extreme case seems particularly appropriate for a qualitative 
description when $c_p > 2$. Here, the first order character of the transition 
in the pure system and the corresponding favored formation of small loops 
suggests that larger differences of local AT to GC content ratio are necessary 
for obtaining different local melting temperatures.

In conclusion, important finite size corrections to scaling are expected 
qualitatively, which could in particular depend strongly on the parameter $x$ 
introduced above, involving both the energy ratio $R$ and the connectivity 
constant $\mu$. Specifically, the effect of disorder on the behavior of the 
system could become evident only for chain lengths diverging exponentially 
with $x$. This parameter seems therefore to play the role of an 
{\em intrinsic} length scale for the {\em rare regions}, corresponding to the 
logarithm of an intrinsic length scale for the system itself.  

Setting aside other differences, the parameter choices in \cite{GaMo,MoGa} 
correspond to $x \sim 15$, whereas in \cite{Co} they give $x \simeq 1.3$. It 
is accordingly possible that results in the two studies can be explained 
following the described picture, on the basis of the underlying finite 
size effects. It is nevertheless to be noted that in \cite{GaMo,MoGa} 
very large sequences, up to $N \sim 2000000$, were considered. The present 
qualitative picture could anyway explain the observations in \cite{Co}, for an 
apparent dependence of $\nu_r$ on $R$, as an increase in $R$ at fixed 
$\log \mu$ amounts to a decrease in $x$. It is possible that the chain 
lengths $N \leq 800$ affordable in the simulations were not large 
enough and that also the value $\nu_r \simeq 1.33$ obtained with 
$R=\infty$ was affected by finite size corrections to scaling, being 
therefore to be interpreted as a lower bound for the (asymptotic) $\nu_r$.
 
\section{Numerical study}
\subsection{Details of the model}
\noindent
We consider the disordered PS model, with $c_p=2.15$, described by the 
canonical partition function:
\begin{equation}
Z^*_{N,\epsilon}=\sum_{\{s_i\}}  
\prod_i e^{s_i (\beta \epsilon_i-\log \mu)} \prod_k
{\frac{1}{(2l_k)^{c_p}}},
\label{Zlast}
\end{equation}
where $\prod_k$ runs over all the loop lengths $l_k>2$, associated with a 
given configuration of the $\{ s_i\}$. Importantly, here and in the following 
we only impose the condition 
$\sum_k (l_k-1) + \sum_i s_i =l_{tot}+n_{tot} \le N$, thus allowing for free 
ends, with separated strands not involved in a loop at one end-extremity of 
the sequence. Such free-ends are not expected to modify the thermodynamic 
limit behavior \cite{KaMuPe1,KaMuPe2,GaMo}, but the high temperature phase of 
the model will consist accordingly of two strands linked only at $i=1$ 
(instead of a large loop, for a system with both extremities required to be in 
the coupled state). This so-called {\em bound-unbound} ($bu$) model 
corresponds more closely to the one studied usually in on-lattice simulations 
\cite{CaCoGr,CaOrSt,BaCaKaMuOrSt,Co}, and it was also adopted in 
\cite{GaMo,MoGa}. We notice that the factor $\mu^{2(N-l_{tot}-n_{tot})}$ 
cancels out when looking at $Z^*=Z/\mu^{2N}$. 

In the present study, we are moreover implicitly adopting the value $\sigma=1$
for the cooperativity factor, where the parameter $\sigma$ gives a measure for 
the barrier to overcome for the initiation of a loop opening. In realistic 
sequence-specific calculations \cite{Bl,Ye,YeBoLa,YeJo,CaMaBl}, one uses 
typically $\sigma=O(10^{-5})$. It is however not clear what choice for 
$\sigma$ is appropriate when $c_p=2.15$ since in experimental/theoretical 
comparisons an exponent $c_p \sim 1.8$ is generally taken and in a recent 
study \cite{BlCa} it was suggested that the values for $\sigma$ and $c_p$ 
should vary in parallel in order to reproduce correctly experimental melting 
curves. We note that small $\sigma$ values could increase corrections to 
scaling, whereas this parameter is not expected to influence the thermodynamic 
properties. 

Disordered PS models can be solved numerically by writing down recursive 
equations for the partition function with a SIMEX scheme 
\cite{FiFr,Yeetal,Ye1,GaOr}, taking into account efficiently the
long-range entropic loop weights. A basic idea in the recursive scheme is that 
the {\em forward} partition function $Z^f_\epsilon(\rho+1)$, which accounts 
for all configurations up to position $\rho+1$ along the sequence 
with both base pairs at positions $i=1$ and $i=\rho+1$ in the closed state 
($s_1=s_{\rho+1}=1$), can be obtained from $Z^f_\epsilon(\rho)$:
\begin{equation}
Z^f_\epsilon(\rho+1)=\exp({\beta \epsilon_{\rho+1}-\log\mu}) \left [
Z^f_\epsilon(\rho)+ \sum_{\rho'=1}^{\rho-1} 
\frac{Z^f_\epsilon(\rho')}{[2(\rho-\rho'+1)]^{c_p}} \right ].
\label{zf}
\end{equation}
We can similarly write down the equation for the {\em backward} partition 
function:
\begin{equation}
Z^b_\epsilon(\rho-1)=\exp({\beta \epsilon_{\rho-1}-\log\mu}) \left [
Z^b_\epsilon(\rho)+ \sum_{\rho'=\rho+1}^{N} 
\frac{Z^b_\epsilon(\rho') }{[2(\rho'-\rho+1)]^{c_p}} + 1 \right ],
\label{zb}
\end{equation}
with the last term in the equation above corresponding to the free-end 
configuration ($s_{i}=0$ for $i>(\rho-1)$).
The canonical partition function for the complete chain of length $N$ is given 
by:
\begin{equation}
Z^*_{N,\epsilon}=\sum_{\rho=1}^{N}Z^f_\epsilon(\rho)=
Z^b_\epsilon(1),
\end{equation}
where the forward sum takes into account the free-ends. With these 
calculations one obtains in particular the probability for a base pair at 
position $i$ (along the sequence of length $N$) to be in the closed state as:
\begin{equation}
{\cal P}_\epsilon(i,T,N)= \langle s_i \rangle = 
\frac{Z^f_\epsilon(i)Z^b_\epsilon(i)}
{Z^*_{N,\epsilon} \exp({\beta \epsilon_i-\log\mu})},
\label{Ps}
\end{equation}
where the division with $\exp({\beta \epsilon_i-\log\mu})$ is to rectify 
the double counting of the corresponding factor (involved
both in $Z^f_\epsilon$ and in $Z^b_\epsilon$).

\subsection{Measured observables}
\noindent
For a given disorder sequence $\{ \epsilon_i, \: i=1,\dots, N \}$, at fixed 
chain length $N$ and temperature $T$, we can derive from 
the ${\cal P}_\epsilon(i,T,N)= \langle s_i \rangle$ (\ref{Ps})  
quantities of interest such as the density of closed AT base pairs 
$\theta_{AT,\epsilon}$ (respectively GC, $\theta_{GC,\epsilon}$), 
the total density of closed base pairs $\theta_{\epsilon}$ and the energy 
density $e_\epsilon$:
\begin{eqnarray}
\theta_{AT,\epsilon}(T,N) & = & \frac{1}{N} 
\left \langle \sum_{i \in AT} s_i \right \rangle=
\frac{1}{N} \sum_{i \in AT} {\cal P}_\epsilon(i,T,N), \\
\theta_{GC,\epsilon}(T,N) & = & \frac{1}{N} 
\left \langle \sum_{i \in GC} s_i \right \rangle=
\frac{1}{N} \sum_{i \in GC} {\cal P}_\epsilon(i,T,N), \\
\theta_\epsilon(T,N)&=&
\theta_{AT,\epsilon}(T,N)+\theta_{GC,\epsilon}(T,N) \\
e_\epsilon(T,N)&=&-\left [ 
\epsilon_{AT} \theta_{AT,\epsilon}(T,N)+ 
\epsilon_{GC} \theta_{GC,\epsilon}(T,N) \right ]. 
\end{eqnarray}
We can also consider the specific heat $c_\epsilon$ as well as the derivative
of the density of opened base pairs $c_{\theta,\epsilon}$, which is relevant
to experimental determinations:
\begin{eqnarray}
c_{\epsilon}(T,N) &=&
\frac{1}{T^2} \frac{de_\epsilon(T,N)}{dT} \\
c_{\theta,{\epsilon}}(T,N)&=&
-\frac{1}{T^2} 
\frac{d\theta_\epsilon(T,N)}{dT}. 
\end{eqnarray}
Since $R\neq 1$, the energy density $e_\epsilon$ and the order parameter 
$\theta_\epsilon$ can exhibit different behaviors, and accordingly such can be 
also the case for $c_{\epsilon}$ and $c_{\theta,{\epsilon}}$. In the same 
direction we consider also the susceptibility, obtained as:
\begin{eqnarray}
\chi_{\epsilon}(T,N)&=&\frac{1}{N}  
\left [ \left \langle \left ( \sum_{i=1}^N s_i \right )^2 
\right \rangle - \left \langle \left ( \sum_{i=1}^N  s_i \right ) 
\right \rangle^2 \right ] = \nonumber \\
&=& \frac{1}{\beta} \left [ 
\frac {d\theta_{AT,\epsilon}(T,N)}{d\epsilon_{AT}}+
\frac {d\theta_{GC,\epsilon}(T,N)}{d\epsilon_{GC}}
+\frac {d\theta_{AT,\epsilon}(T,N)}{d\epsilon_{GC}}+
\frac {d\theta_{GC,\epsilon}(T,N)}{d\epsilon_{AT}}
\right ],
\label{chi}
\end{eqnarray}
providing interestingly a possibility for checking numerical accuracy in the 
computations, from the fulfillment of the equality 
${d\theta_{AT,\epsilon}(T,N)}/d\epsilon_{GC}=
{d\theta_{GC,\epsilon}(T,N)}/d\epsilon_{AT}$.

We study  moreover the behavior of the (non-normalized) loop 
length probability distribution:
\begin{equation}
P_\epsilon(l,T,N)= {\cal N}(l)\sum_{i=1}^{N-l-1}
\frac{Z^f_\epsilon(i)Z^b_\epsilon(i+l+1)}
{Z^*_{N,\epsilon}}
\end{equation}
with ${\cal N}(l)=1/(2l)^{c_p}$, independent from the disorder sequence.
Therefore we introduce the quantity:
\begin{equation}
P^*_{\epsilon}(l,T,N)=(2l)^{c_p} P_\epsilon(l,T,N).
\label{pstar}
\end{equation}
noting that in the pure model $P^*(l,T,N) \propto \exp({-l/\xi(T,N)})$
(see (\ref{plt})) and correspondingly $P^*(l,T,N) \rightarrow const$ 
for $T\rightarrow T_c^-$.

As a first step in the analysis of such data, we will check the validity
of standard finite size scaling for quantities averaged over disorder. 
With the usual definition of the critical exponents and 
$y=(T-T_c)N^{1/\nu_r}$, one expects:
\begin{eqnarray}
\overline{e_\epsilon(T,N)}& \sim& N^{1/\nu_r-1} 
\tilde{e}(y)
\label{scle}\\
\overline{c_\epsilon(T_c,N)}& \sim & N^{2/\nu_r-1}
\tilde{c}(y),
\label{maxlaw}\\
\overline{\theta_\epsilon(T,N)} &\sim& N^{-\beta_r/\nu_r}
\tilde{\theta}(y)
\label{sclop}\\
\overline{\chi_\epsilon(T,N)} &\sim& N^{\gamma_r/\nu_r}
\tilde{\chi}(y),
\label{sclsu}
\end{eqnarray}
where it is possible to find $\beta_r/\nu_r \neq 1-1/\nu_r$ and 
$\gamma_r/\nu_r \neq 2/\nu_r-1=\alpha_r/\nu_r$.
For the average loop length probability distribution, we still look for  
a behavior at the critical point described by a power law as in the pure
case: 
\begin{equation}
\overline{P_\epsilon(l,T_c,N)} \propto 1/l^{c_r},
\label{plpl}
\end{equation}
and therefore
\begin{equation}
\overline{P^*_\epsilon(l,T_c,N)} \propto l^{c_p-c_r}.
\label{pstartc}
\end{equation}
Interestingly, based on this relation, it should be particularly simple 
to seek numerical evidence for $c_r \neq c_p$.
 
\subsection{Computational details}
\noindent
We resort to the SIMEX scheme \cite{Yeetal,Ye1,GaOr}, based explicitly on 
partition function evaluations, instead of recursions for specific 
conditional probabilities as in \cite{FiFr}. Besides this conceptual 
difference (important for generalizations, notably to higher-order models), 
for the simple helix-coil model in linear molecules, as considered here, the 
reduction of the computational complexity by one order of magnitude 
in the SIMEX method relies on 
the numerical representation of the long-range effects in the model as a sum 
of $N_{S}$ exponentials, as already formulated in the FF method.
The other important ingredient in the FF, also implemented straightforwardly
in the SIMEX, corresponds to a forward-backward scheme as described 
in Section 3.1, classical in dynamic programming and associated with 
an additional order of magnitude reduction in complexities. For the linear 
case, the complexities for a complete probability map 
calculation reduce overall from 
$N^3$ (for a one-way progressive treatment) to $N_{S}N$.

In order to make the scheme operational in practice, it is necessary to 
obtain appropriate numerical representations for long-range effects 
as sums of exponentials. The general numerical problem associated with the 
analysis of multiexponential functions is notoriously a delicate one. It 
covers two distinct -in principle- situations, concerning either 
identifications or approximations, the relevant case for the present study. In 
the identification situation, it is necessary to recover the correct number of 
exponentials, and of course the correct associated parameters, from curves 
(usually experimental) supposed to be of multiexponential type for theoretical 
reasons. A general solution to this problem is provided by the Pad\'e-Laplace 
method \cite{YeCl}, requiring no {\em a priori} hypotheses for the 
identification of components in sums of general exponentials (real and/or 
complex). This formulation encompasses, and generalizes, in a unified frame, 
a series of solutions since Prony's method and the so-called method of moments 
\cite{MeMo}. Even though originally formulated as an identification approach, 
the numerical application of the Pad\'e-Laplace method to power-law functions 
(such as for loop-entropies here) revealed an identification-like behavior in 
this approximation problem, in the sense that, for given maximal long-range 
lengths, a fixed number of significant exponential components are obtained. 
For example, for a series of biologically-oriented studies, an approximation 
of $1/l^{1.95}$ with $N_S=14$ exponentials was shown to be appropriate (with 
further refinements of the parameters with least-squares procedures) 
\cite{Ye}. In the present study, in order to be in strictly comparable 
conditions with this respect, we adopt the numerical representation of 
$1/l^{2.15}$ with $N_S=15$ exponentials in \cite{GaMo}. 

For the numerical computation of the recursive equations for the forward and 
backward partition functions, it is moreover necessary to avoid 
underflow/overflow problems. For this purpose different schemes can be 
implemented \cite{Yeetal,GaOr}, in order to normalize the numerators and 
denominators the ratios of which are involved in the evaluation of 
probabilities (\ref{Ps}). We consider here the normalization described in 
\cite{GaOr}, based on the introduction of {\em free energy like quantities} 
for the handling of the logarithms of $Z^f_\epsilon$ and $Z^b_\epsilon$. 
The details of the implementations are provided in the Appendix, along with 
the description of boundary conditions.

We study extensively the case $x=R/[\log \mu (R-1)]=1.3$, using the same 
energies and connectivity constant as in \cite{Co}: $R=2$ and 
$\log \mu \simeq \log \mu_{SAW} \simeq 1.54$, in three dimensions. We consider 
sequence lengths ranging between $N=100$ and $N=20000$. For $x=1.3$, such 
$N$-values appear to be large enough both for the clarification of the 
thermodynamic limit behavior and for the study of corrections to scaling. 
On the other hand, the numerical computations are reasonably fast up to this 
length, which makes it possible to consider closely spaced temperatures
and to obtain correspondingly, with negligible numerical errors, quantities
related to derivatives, such as in particular the values of the maxima
of the specific heat. In detail, for given chain lengths, we consider 
${\cal N}_T=250$ different $T$-values, equally spaced in intervals 
$[T_{min}(N),T_{max}(N)]$ around the corresponding $T_c(N)$, evaluated roughly 
from the position of the maximum of the average specific heat in some 
preliminary results. For the different chain lengths, the number of samples 
${\cal N}_s(N)$ as well as the $T_{min}(N)$ and $T_{max}(N)$ temperatures are 
detailed in [Tab. 1]. 

\begin{table}
\begin{center}
\begin{tabular}{||c|c|c|c||}
\hline
\hline
$N$ & ${\cal N}_s(N)$ & $T_{min}(N)$ &$T_{max}(N)$ \\
\hline
\hline
100 & 2000 & 0.95 & 1.2 \\
\hline
200 & 2000 & 0.95 & 1.2 \\
\hline
500 & 2000 & 1.0 & 1.16 \\
\hline
750 & 1000 & 1.0 & 1.16 \\
\hline
1000 & 1000 & 1.0 & 1.15 \\
\hline
2500 & 1000 & 1.02 & 1.14 \\
\hline
5000 & 1000 & 1.02 & 1.14 \\
\hline
7500 & 1000 & 1.04 & 1.12 \\
\hline
10000 & 600 & 1.04 & 1.14 \\
\hline
15000 & 500 & 1.04 & 1.12 \\
\hline
20000 & 500 & 1.04 & 1.12 \\
\hline
\hline
\end{tabular}
\caption{Tabulation of the chain lengths $N$, the number of samples 
${\cal N}_s(N)$ and the range of temperatures $[T_{min}(N),T_{max}(N)]$ in the 
numerical computations. For each disordered sequence ${\cal N}_T=250$ equally 
spaced temperatures, in the corresponding intervals, are considered.}
\end{center}
\end{table}

It was checked in particular that with the various choices above the errors on 
the maximum of specific heat and of susceptibility were both consistently 
smaller than fluctuations between samples. Without any loss of generality we 
set in all calculations $\epsilon_{AT}=1$, i.e. temperature is in 
$\epsilon_{AT}$ unities. The evaluation of the susceptibility is obtained by 
numerical derivations with respect to $\epsilon_{AT}$ and $\epsilon_{GC}$ 
(see (\ref{chi})), with $\delta \epsilon_{AT}=\beta \cdot 10^{-4}$ and
$\delta \epsilon_{GC}=R \delta \epsilon_{AT}$, which ensures the desired 
numerical accuracy at all temperatures. Finally, in all calculations the 
errors on average quantities are computed from sample-to-sample fluctuations.

\section{Results and discussion}
\subsection{Given sample-sequence and different $x$ values}
\noindent
We consider first the qualitative behavior of the model for a given 
sample-sequence of length $N=10000$, with different $x=R/[\log \mu (R-1)]$ 
values. Results shown in this Section are obtained with a particular disorder 
configuration. It was however checked that the corresponding qualitative 
observations are also valid with various arbitrarily chosen sequences. 
In order to cover different significant situations, the following 
values for the parameter $x$ were considered:  
$x = 2, \: 1.3, \: 1.4, \: 0.7$, respectively. Notably, the choice $x=2$ is 
for comparisons with the results in \cite{GaMo,MoGa}. In detail, we used the 
same $R=1.098$, and we set in addition $\log \mu \simeq 5.55$, for obtaining 
close critical temperatures in the pure case. The choice $x=1.3$ is for 
compatibility with the conditions in \cite{Co}, and accordingly we set in this 
case $R=2$ and $\log \mu \simeq 1.54$. On the other hand, the choice of the 
close value $x \simeq 1.4$ is following parameters setting usual in 
comparisons with experimental results \cite{Bl,Ye,YeBoLa,YeJo}. In this latter 
case, the value for $x$ is not related to a large $R$ value, but rather to a 
large average $\log \mu \simeq 12.35$ (in $k_B$ unities). It can be noticed 
that typically used coupling energies lead to $R \simeq 1.062$, as obtained by 
averaging over the different stacking energies for neighbor base pairs 
$\epsilon_{i,i+1}$. Finally, for clarification of potential differences 
resulting from choices of large $R$ or alternatively large $\mu$ values, 
$x \simeq 0.7$ was retained as corresponding to the two choices for 
($R$, $\log \mu$) couples: $R\simeq 1.062$, $\log \mu \simeq 24.5$ (case $a$); 
and $R \simeq 18.38$, $\log \mu \simeq 1.54$ (case $b$).

\begin{figure}[hpbt]
\begin{center}
\leavevmode
\includegraphics*[angle=270,width=8cm]{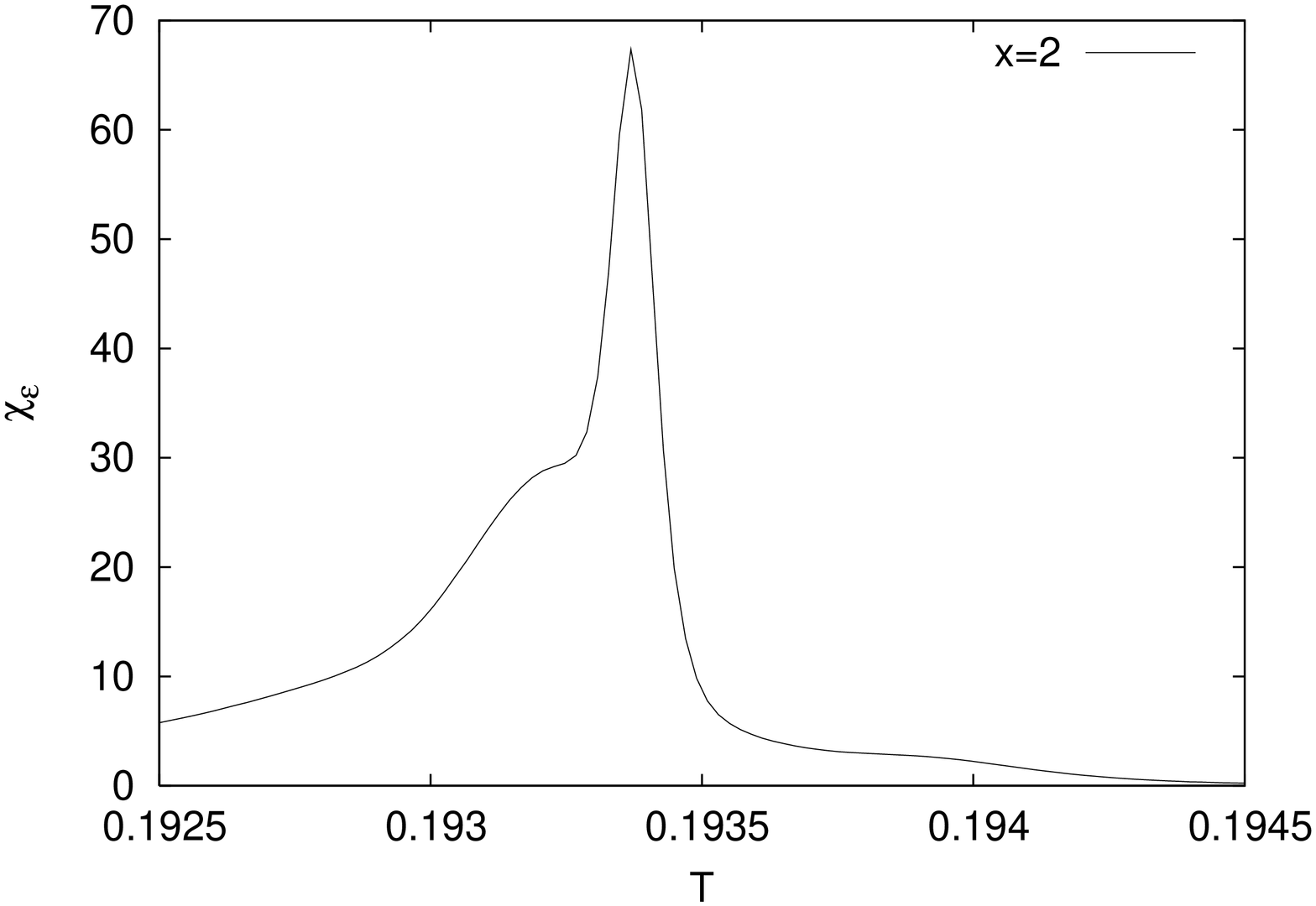}\\
\includegraphics*[angle=270,width=8cm]{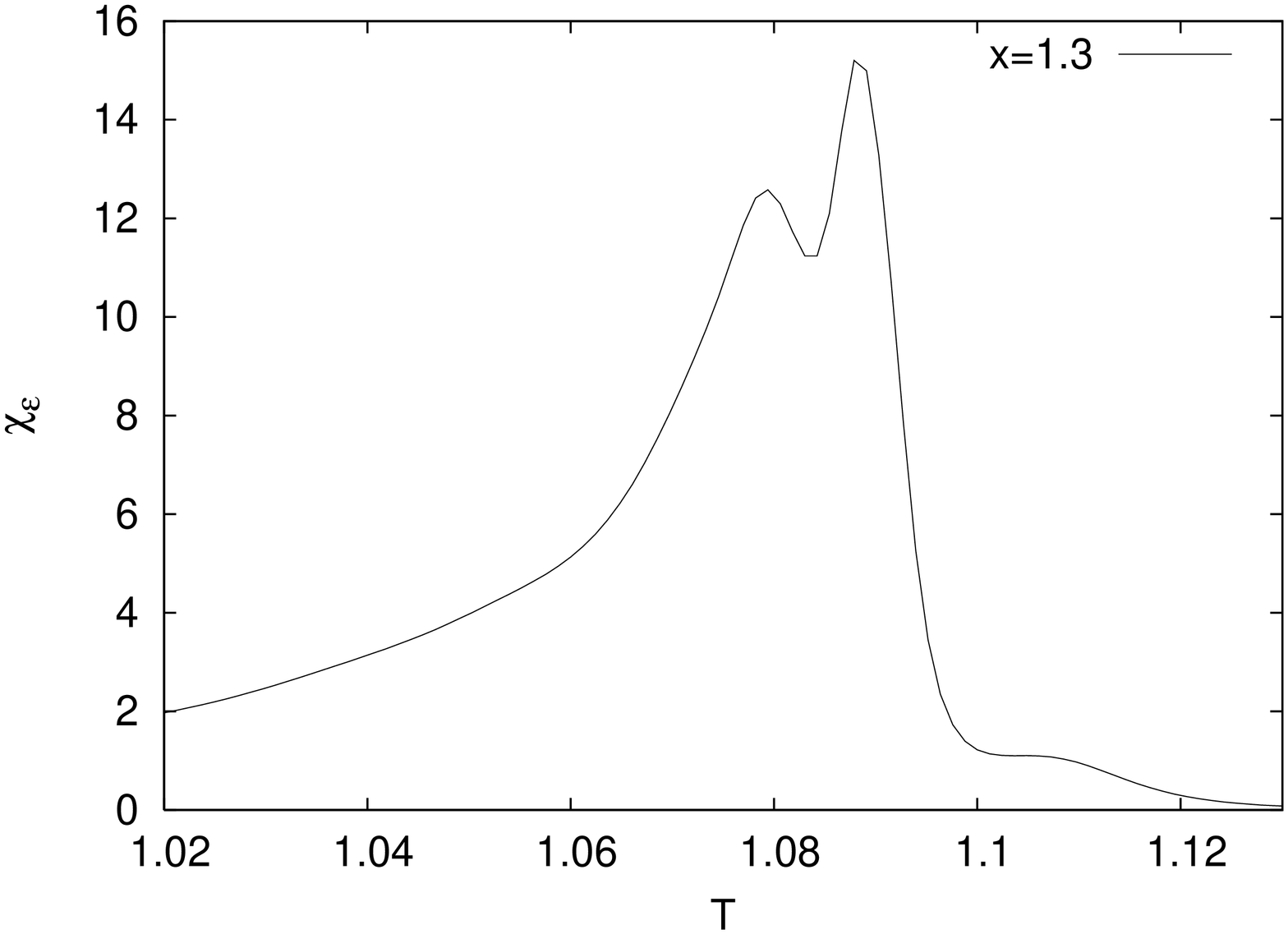}
\includegraphics*[angle=270,width=8cm]{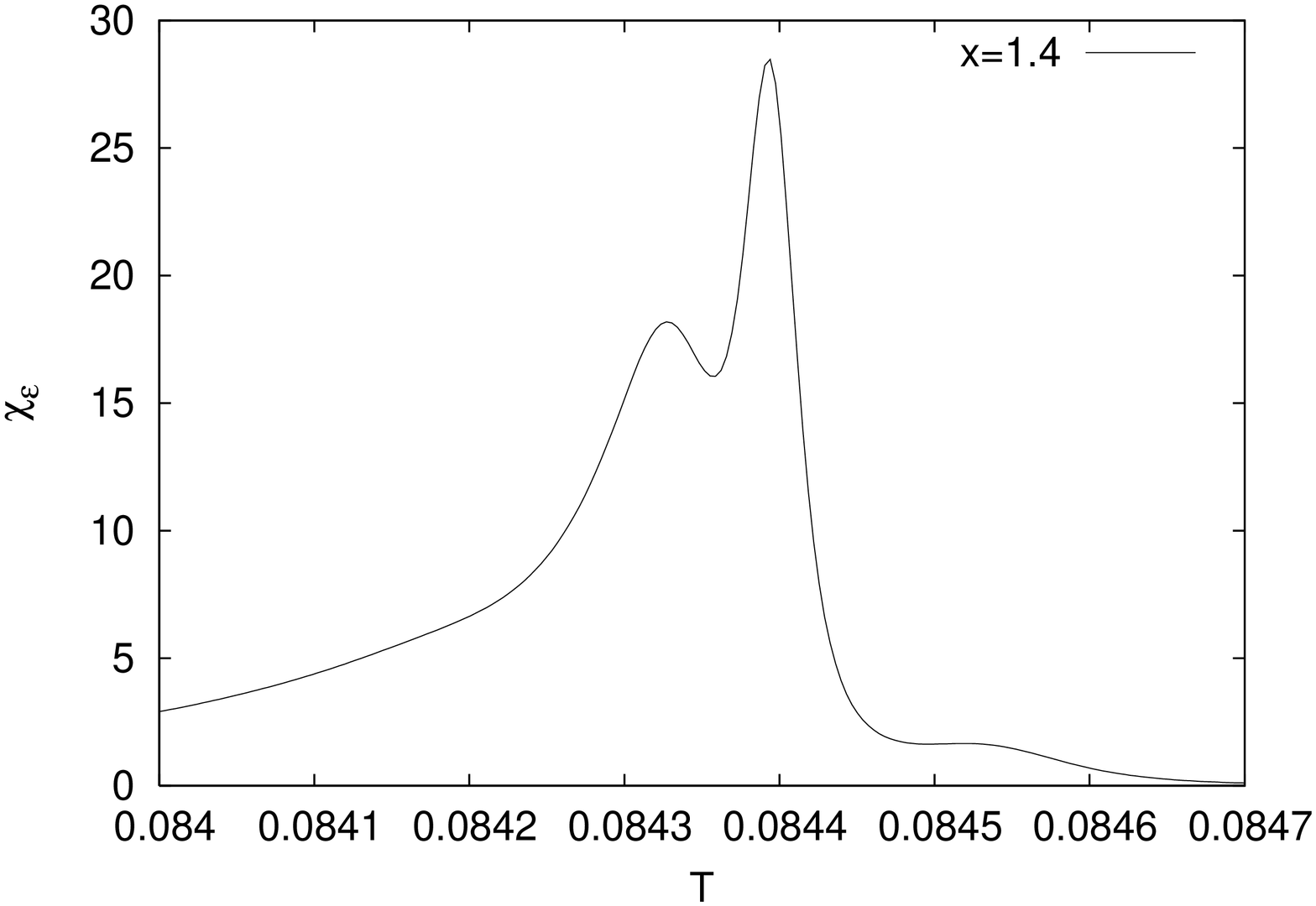}
\includegraphics*[angle=270,width=8cm]{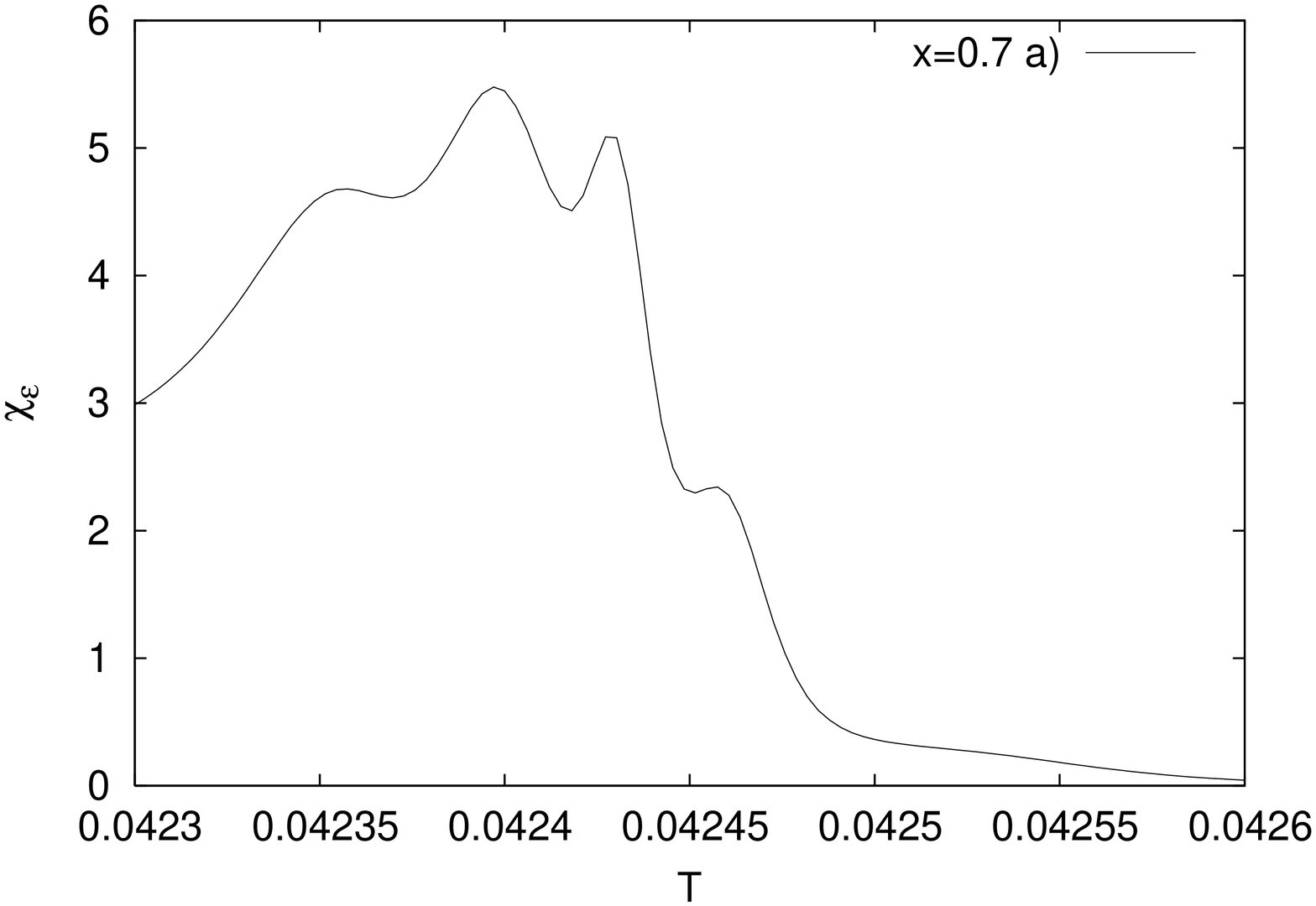}
\includegraphics*[angle=270,width=8cm]{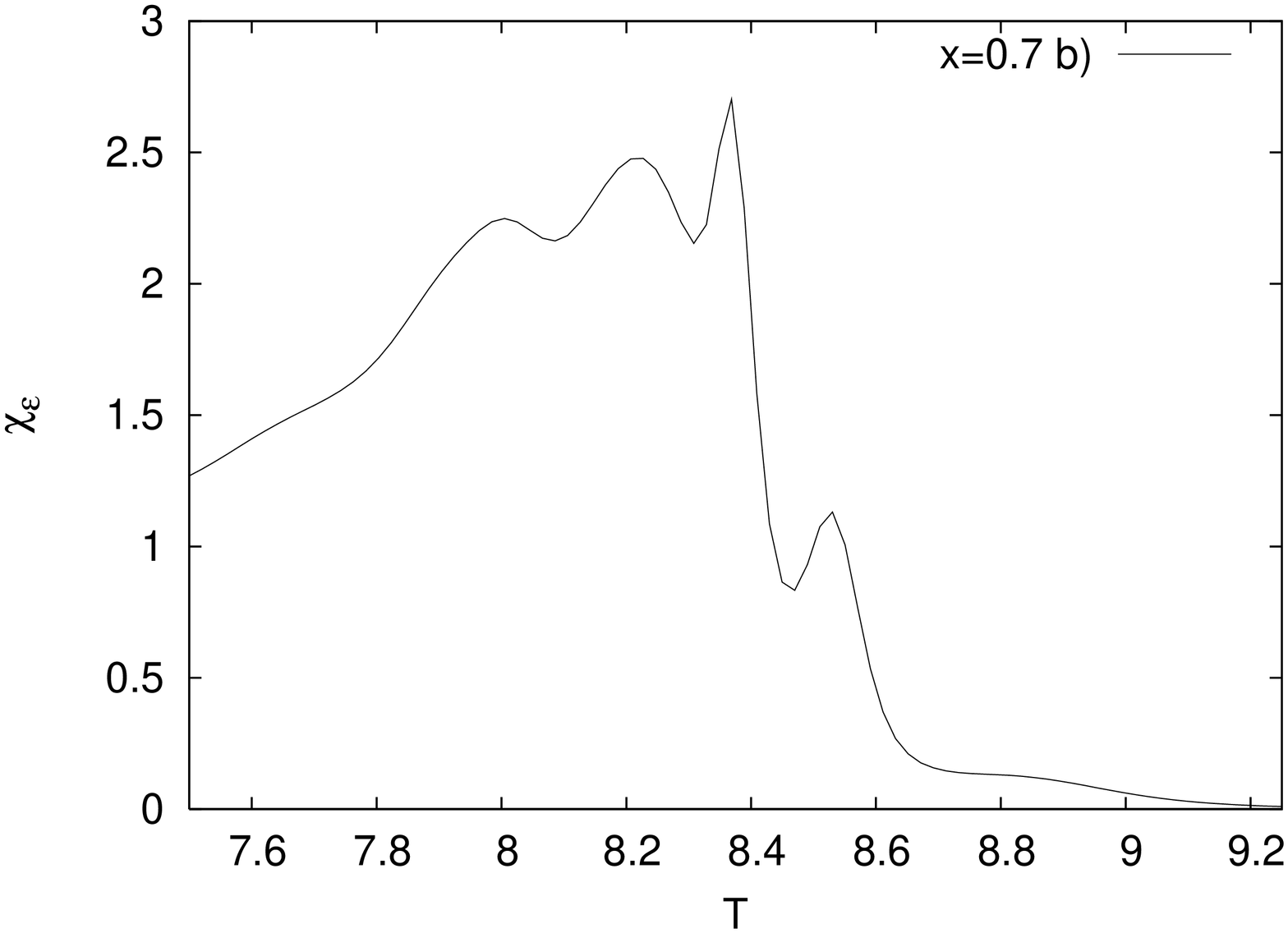}
\caption{Susceptibility curves for a given sample-sequence of length
$N=10000$ and different $x$ values (see text).}
\label{fig1}
\end{center}
\end{figure}

We plot in [Fig. \ref{fig1}] the susceptibilities $\chi_\epsilon(T,x,N)$ for 
the sample-sequence for the different $x$ values. We observe that the results 
depend strongly on $x$. However, the shapes of the curves obtained with 
$x=1.3$ and $x = 1.4$ appear to be strikingly similar. Moreover, also the two 
curves corresponding to the two different cases associated with $x = 0.7$ 
(case $a$ and case $b$) are qualitatively similar.  
\begin{figure}[hpbt]
\begin{center}
\leavevmode
\includegraphics*[angle=270,width=10cm]{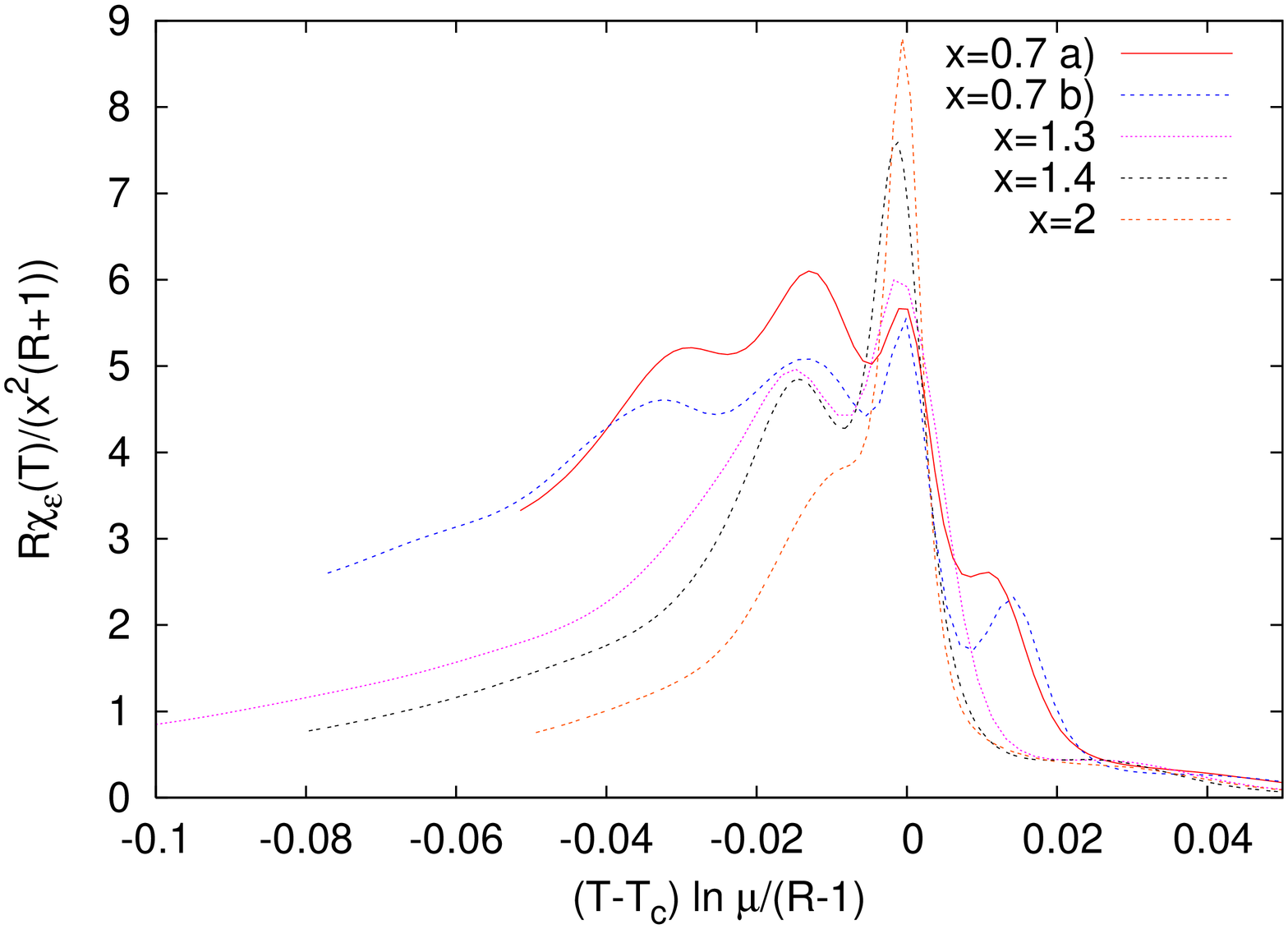}
\caption{Scaled susceptibility for a given sample-sequence of length
$N=10000$ and different $x$ values. $\chi_\epsilon R/ [x^2 (R+1)]$ data are 
plotted as function of $[T-T_c(\epsilon,x,N)] \log \mu/(R-1)$, with 
$T_c(\epsilon,x,N)$ the temperature associated with the absolute maximum 
of susceptibility. In the case $a$ with the value $x= 0.7$,
specific choice was needed for defining the pseudo-$T_c$,
and we looked in this case to correspondence of positions of the peaks with 
the other curves, by taking it as the temperature of the second maximum.}
\label{fig2}
\end{center}
\end{figure}

The results here are in agreement with the overall picture given in Section 
2.4, with indications for an $x$-dependent finite size behavior. The extreme 
case considered there, with pure AT and GC regions, is clearly a very rough 
approximation of a typical sequence. It is nevertheless clear 
that the parameter $x$ appears to capture some essential ingredients
of the model.

For more quantitative analysis, we note that for $N \rightarrow \infty$ one 
expects a transition temperature close to $T_c^{\infty} \sim
(T_{c,{AT}}+T_{c,{GC}})/2 \sim (R+1)/(2\log \mu)$. For a given finite 
sequence, a pseudo-critical temperature $T_c(\epsilon,x,N)$ must 
be adopted, as the critical temperature is not well defined. Here we take 
as $T_c(\epsilon,x,N)$ the temperature associated with the highest maximum 
value for susceptibility \cite{WiDo}. In particular, with such a choice, we 
find that the behavior of $\chi_\epsilon(T,x,N)$ displays some scaling when 
plotted as function of $[T-T_c(\epsilon,x,N)] T_c(\epsilon,x,N)/(R-1)$, 
for different $R$ and $x$ values. We present in [Fig. \ref{fig2}] 
correspondingly scaled $\chi_\epsilon$ data (also multiplied by the factor 
$R/[x^2(R+1)]$, for obtaining close behaviors in the high temperature phase).

This figure further makes evident the dependence on $x$ for finite size 
systems. The results suggest that, at fixed length scale $N$, for large $x$, 
and in particular here already for the value $x=2$, the system exhibits 
typically only one very sharp peak. This observation should be related to the 
fact that the probability of large enough {\em rare regions} is negligible 
(though one could still encounter such a case when considering a large number 
of sequences), and the system behaves essentially as a pure model with 
$\epsilon=(\epsilon_{AT}+\epsilon_{GC})/2=(R+1)/2$. For smaller $x$ values, we 
observe on the contrary an increasing number of peaks, with decreased 
sharpness. This finding is coherent with the qualitative picture following 
which, with increasingly smaller $x$ values, {\em rare regions} with 
increasingly smaller lengths $L$ are sufficient in order to observe multistep 
behaviors, since the relevant quantity should be the ratio $L/x$. 
Nevertheless, for the smallest considered value $x=0.7$, obtained with two 
different parameters choices, we observe the same number (four) of peaks, but 
the position with respect to the other peaks for the absolute maximum of the 
curve is shifted for $x=0.7 \: a$ as compared to the corresponding one for 
$x=0.7 \: b$ and the larger $x$ values. This confirms that the introduced 
$x$ describes the finite size behavior only approximately.

The changes in the behavior of a typical sample, as described above, are also 
expected at larger sequence sizes for a given $x$-value, since this parameter 
appears to behave as (the logarithm of) an intrinsic length scale of the 
system. Some numerical evidence in this direction was already given in 
\cite{Co} for the on-lattice DSAW-DNA, where the qualitative analysis of the 
specific heat suggested the appearance of multi-peaked curves only for large 
enough $N$-values, and an increasing number of peaks in typical sequences as 
function of $N$. Here we obtain the same qualitative results for the value 
$x=1.3$ studied in detail. Nevertheless, detailed quantitative bases for such 
conclusions are left for future work, with an extensive study of the finite 
size effects for different $x$ values. 

It is to be emphasized that in the suggested picture it is the behavior of the 
{\em typical} finite size sample which is expected to change when varying $x$, 
and therefore one would not expect different {\em typical} and {\em average} 
correlation lengths in the thermodynamic limit, though the sizes necessary for 
confirming this hypothesis could be out of reach for large $x$ values. An 
analysis in terms of pseudo-critical temperatures is clearly necessary to
distinguish between this situation and the alternative one proposed in 
\cite{GaMo,MoGa,Mopr,MoGapr}, which is left to a forthcoming work \cite{CoYe2}.

\subsection{Behavior of average quantities}
\begin{figure}[hptb]
\begin{center}
\leavevmode
\includegraphics*[angle=270,width=8cm]{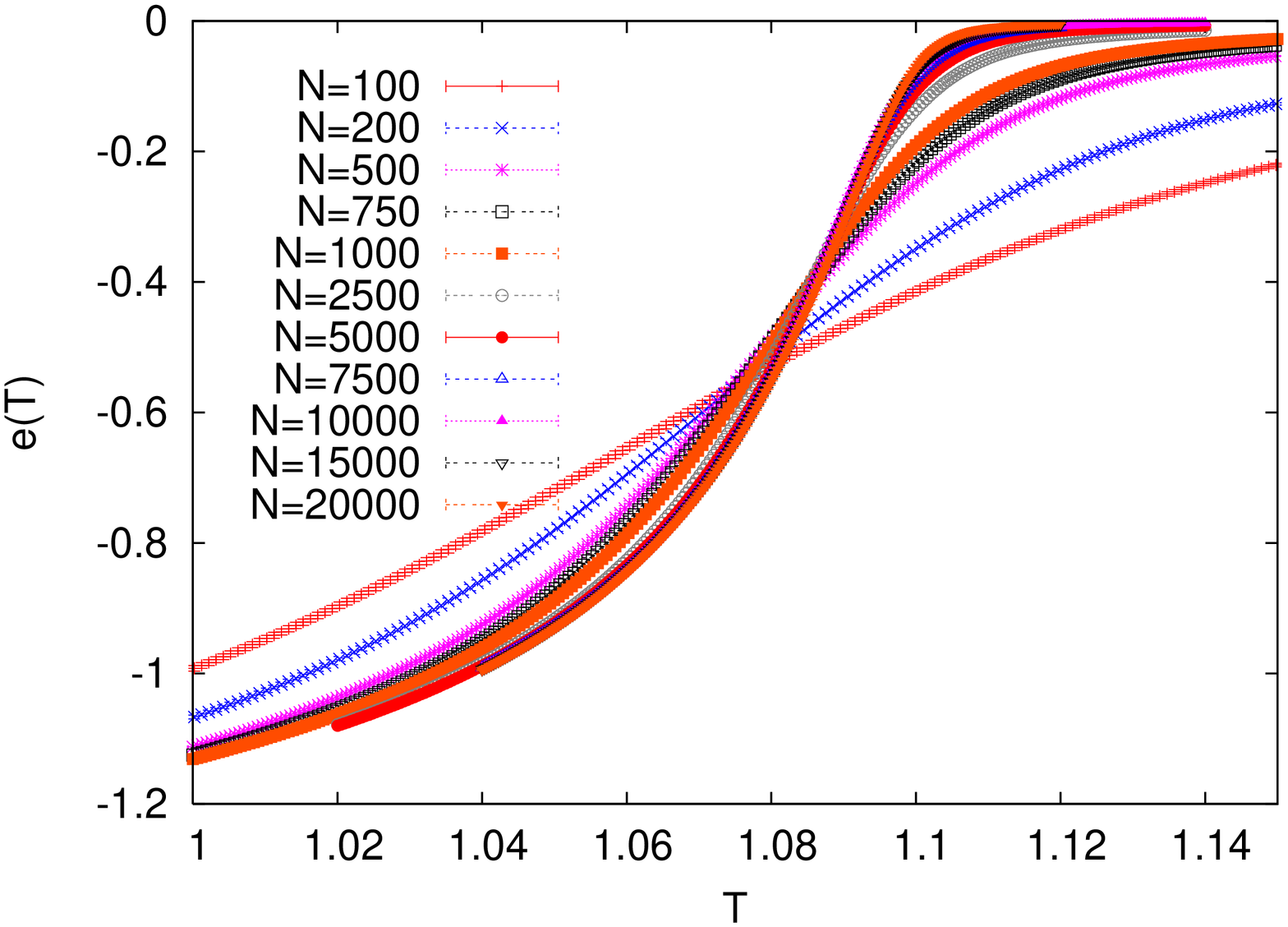}
\includegraphics*[angle=270,width=8cm]{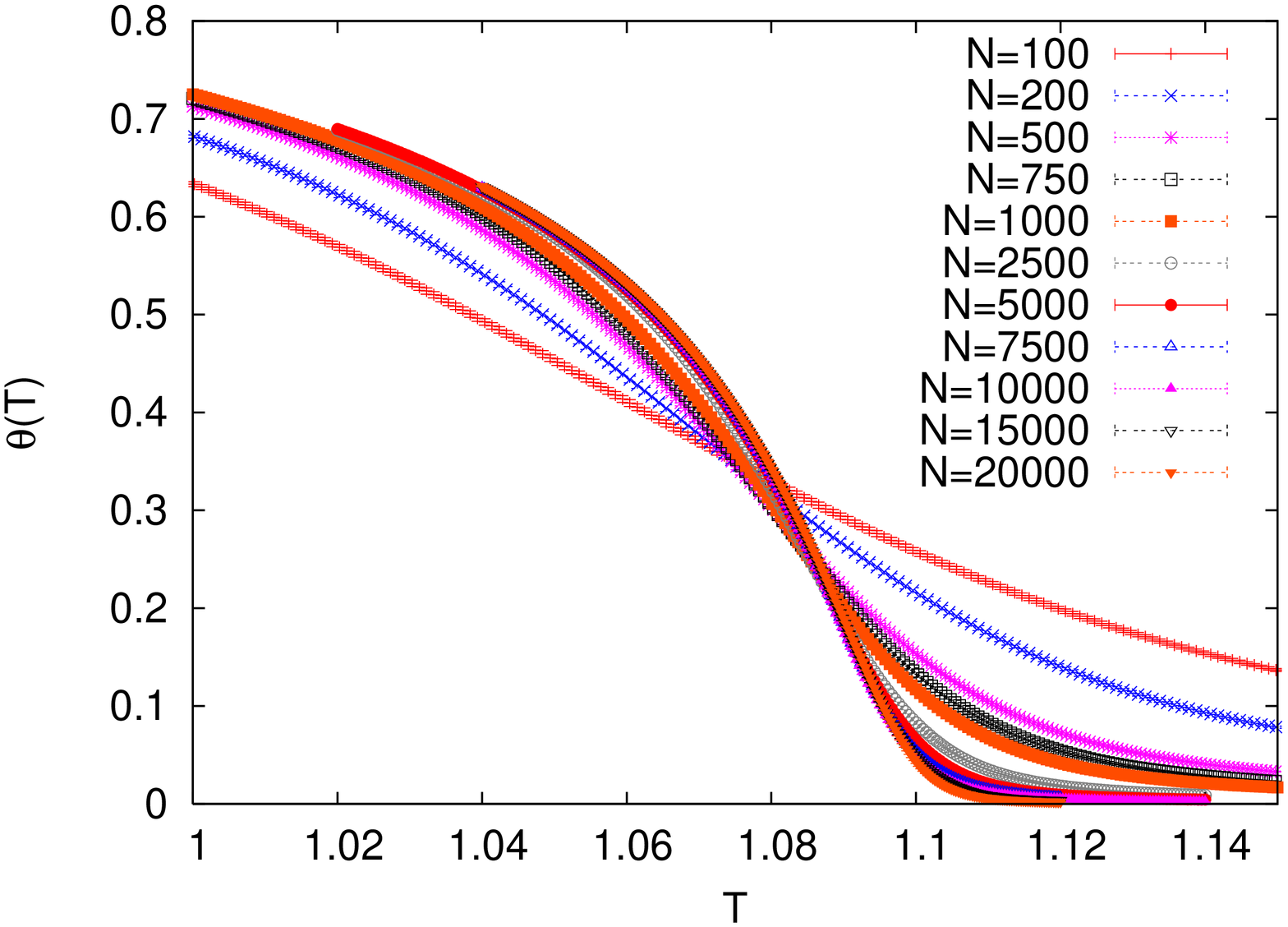}
\caption{Averaged energy density $\overline{e_\epsilon(T,N)}$
and averaged total density of closed base pairs 
$\overline{\theta_\epsilon(T,N)}$, for various chain
lengths, plotted as functions of temperature.}
\label{fig3}
\end{center}
\end{figure}

\noindent
In what follows, we investigate the behavior of average quantities for $x=1.3$ 
(obtained by setting $R=2$ and $\log \mu=1.54$). The average energy density 
$\overline{e_\epsilon(T,N)}$ and the average closed base pair density 
$\overline{\theta_\epsilon(T,N)}$, for different chain lengths, are 
represented in [Fig. \ref{fig3}]. It is clear that the system undergoes a 
phase transition in the thermodynamic limit, with the energy density and the 
order parameter going from zero at high temperature to finite values below 
$T_c$. We can moreover observe from [Fig. \ref{fig3}] very similar behaviors 
for the two quantities (apart from the sign difference) and we expect to 
find, as a consequence, $\beta_r=\nu_r-1$ and $\gamma_r=\alpha_r=2-\nu_r$. 
We have also checked that the average densities of closed AT 
$\overline{\theta_{AT,\epsilon}(T,N)}$ and GC 
$\overline{\theta_{GC,\epsilon}(T,N)}$ base pairs exhibit 
qualitatively similar behaviors. 

Nevertheless, the data do not agree with the corresponding expected scaling 
laws (see (\ref{scle}) and (\ref{sclop})), making clear the presence of strong 
corrections to the asymptotic behavior. Even though it is therefore difficult 
to evaluate the critical exponents, the fact that the curves do not cross at 
the same point (as particularly evident for the largest sizes) suggests a 
transition with (an average) $\nu_r>1 \: (=\nu_p)$. More in detail, the energy 
density and the order parameter appear to converge both towards functions 
which are continuously vanishing at the critical point and possibly also 
differentiable, which would imply a transition at least of second order.

\begin{figure}[hptb]
\begin{center}
\leavevmode
\includegraphics*[angle=270,width=10cm]{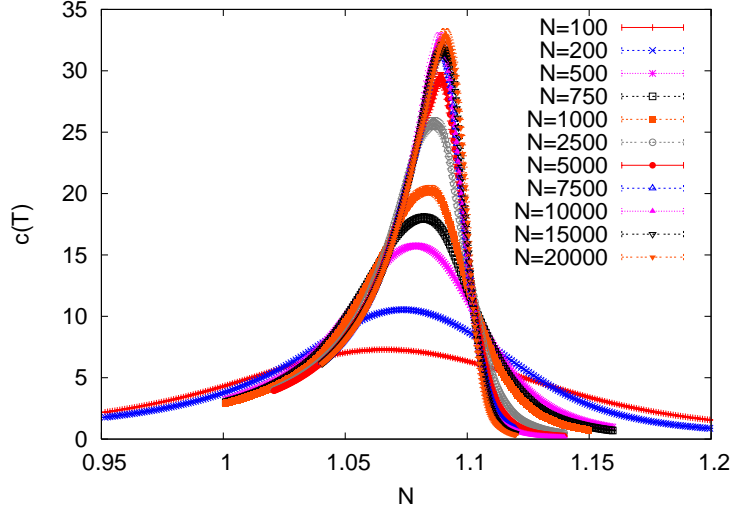}
\caption{Average specific heats $\overline{c_\epsilon(T,N)}$, 
for the chain lengths considered, plotted as functions of temperature.}
\label{fig4}
\end{center}
\end{figure}

Still better evidence for a smooth transition comes from the average specific 
heat $\overline{c_\epsilon(T,N)}$ data, plotted in [Fig. \ref{fig4}]. One can 
notice from this figure that the maximum appears to diverge with the chain 
length for the smallest sizes, but saturates for larger $N$-values, as 
expected for a critical point characterized by $\alpha_r \le 0$, {\em i.e.} 
$\nu_r \ge 2/d \:(=2)$. Interestingly, the qualitative behavior for chain 
lengths smaller than $\sim 1000$ appears to be similar to the one found in 
\cite{Co}. This observation suggests that the on-lattice DSAW-DNA and the 
off-lattice disordered PS model considered could exhibit the same kind of 
finite size effects, also supporting the hypothesis that the value 
$\nu_r \sim 1.2$ obtained from the Monte Carlo like numerical simulations 
represents a lower bound for the average correlation length critical exponent.

\begin{figure}[hptb]
\begin{center}
\leavevmode
\includegraphics*[angle=270,width=10cm]{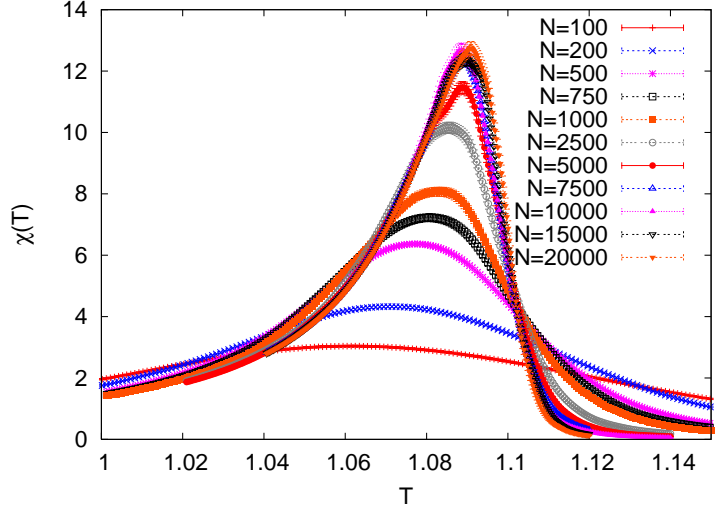}
\caption{Averaged susceptibilities $\overline{\chi_\epsilon(T,N)}$, 
for the chain lengths considered, plotted as functions of temperature.}
\label{fig5}
\end{center}
\end{figure}

\begin{figure}[hptb]
\begin{center}
\leavevmode
\includegraphics*[angle=270,width=10cm]{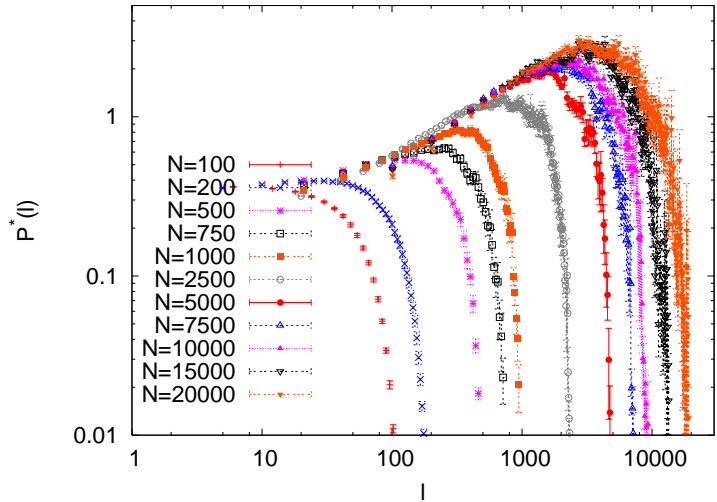}
\caption{Average $\overline{P^*_\epsilon(l,T_c(N),N)}=(2l)^{c_p} 
\overline{P_\epsilon(l,T_c(N),N)}$, for the chain lengths considered, plotted 
as functions of the loop lengths $l$ at temperatures $T_c(N)$, corresponding 
to maxima of average specific heat. Normalization factors are arbitrarily 
chosen in order to make values comparable at small $l$.}
\label{fig6}
\end{center}
\end{figure}

We find a very similar behavior for the susceptibility 
$\overline{\chi_\epsilon(T,N)}$, represented in [Fig. \ref{fig5}], as well as 
for the derivative with respect to the temperature of the total density of 
closed base pairs $\overline{c_{\theta,\epsilon}(T,N)}$ (not shown). These 
findings further suggest that these quantities, as well as the energy and the 
order parameter, are described also in the disordered case by the same 
critical exponents. Again, it is difficult to evaluate the exponents by 
applying standard finite size scaling analysis to the data, because of the 
obvious presence of strong corrections to the expected laws 
(here (\ref{maxlaw}) and (\ref{sclsu})). 

Finally, [Fig. \ref{fig6}] displays data for 
$\overline{P^*_\epsilon(l,T_c(N),N)}$ (see (\ref{pstar})) at the 
size-dependent critical temperatures $T_c(N)$, identified here with the 
temperatures associated with the maximum of the average specific heat. We 
note also in this case a strongly $N$-dependent behavior. In particular,
the considered quantity is nearly constant in the range $1 \ll l \ll N$ for 
the smallest sizes, whereas for the largest ones, expected to be the most 
meaningful, it is increasing with $l$ rather linearly (on logarithmic scale) 
on a wide ranged interval, which should mean that the expected power law 
representation (\ref{plpl}) is valid, but with $c_r \ll c_p$. 

We notice that the observed strong $N$-dependence of averaged quantities 
and the fact that data do not obey usual scaling laws on the whole $N$-range 
studied is in agreement with the qualitative picture given in
Section 2.3, clearly suggesting that, at fixed $x$, the effect of disorder
becomes evident, and the system reaches the asymptotic behavior, only
for large enough size values. From this point of view, in particular the
saturation of the specific heat and susceptibility maxima can be related to 
the appearance of a larger number of less sharp peaks (apparently in the 
typical sequences) for increasing $N$, in analogy with the behavior discussed 
in the previous Section when decreasing $x$ at fixed chain length.

\subsection{Critical exponents and corrections to scaling}
\begin{figure}[hptb]
\begin{center}
\leavevmode
\includegraphics*[angle=270,width=8cm]{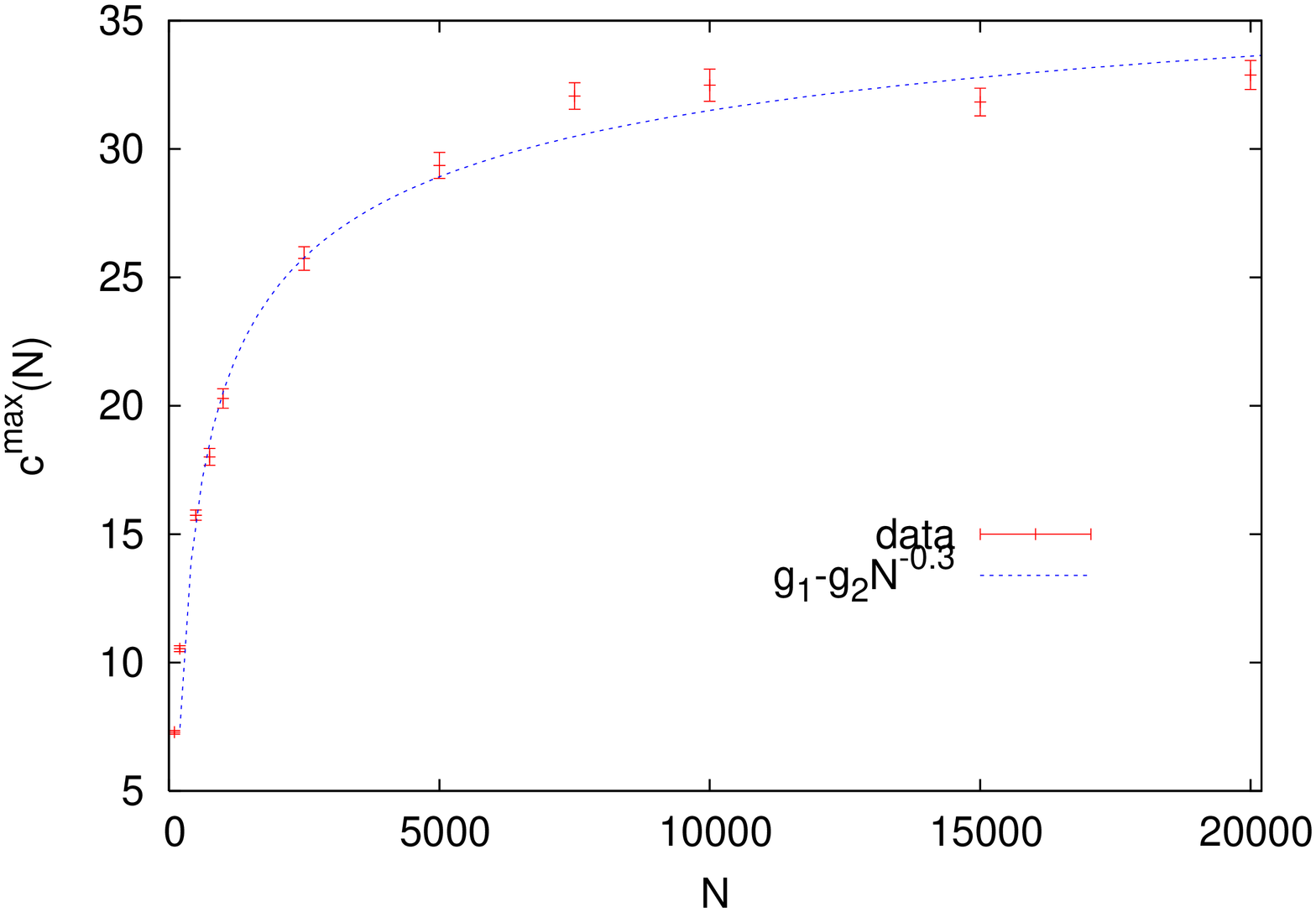}
\includegraphics*[angle=270,width=8cm]{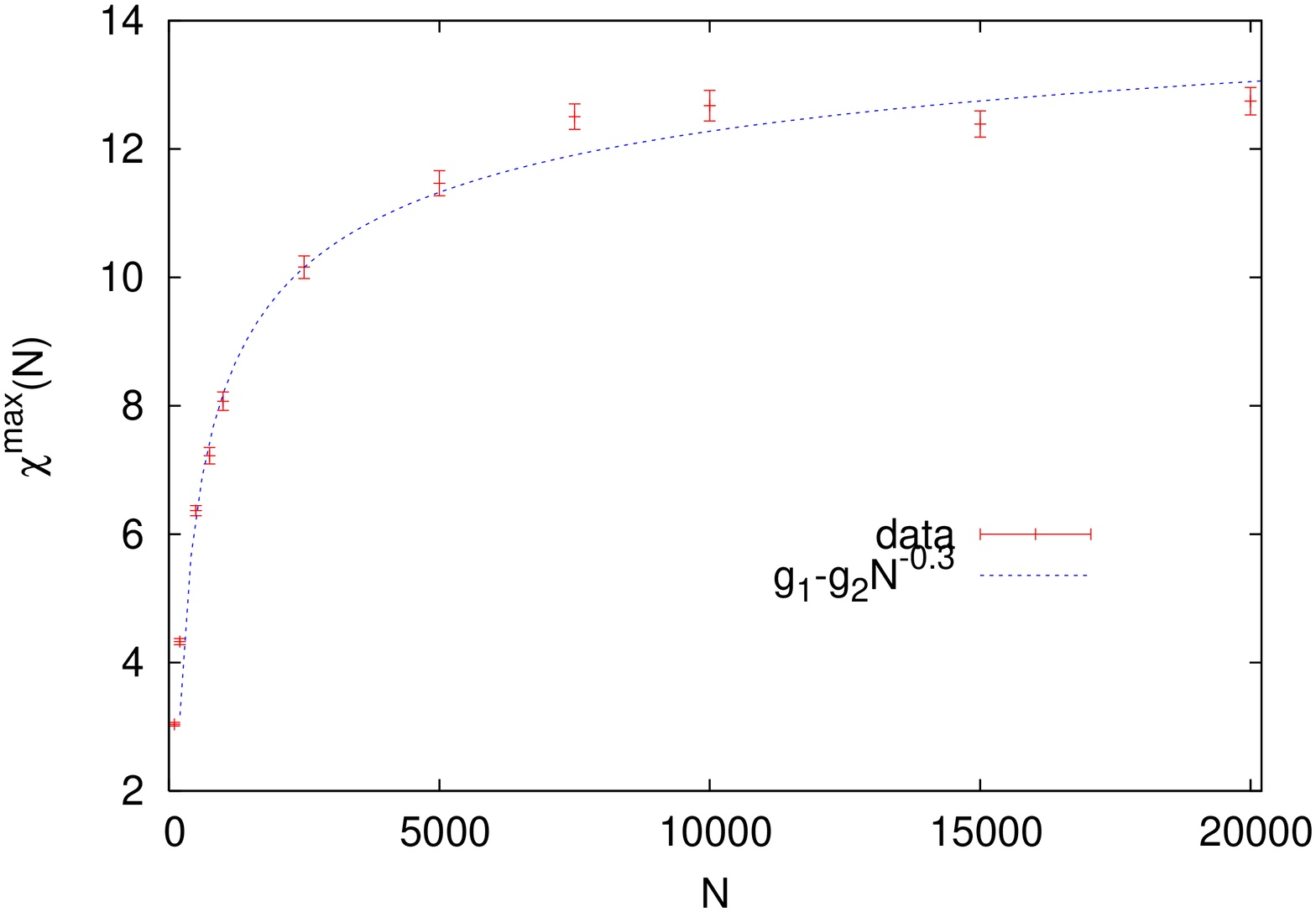}
\caption{Data on the maximum of the average specific heat 
$\overline{c(T,N)}^{max}$ and on the maximum of the average 
susceptibility $\overline{\chi(T,N)}^{max}$, plotted as functions of  
chain lengths, together with the best fits to the law $g_1-g_2N^e$ 
with $g_1,g_2>0$. } 
\label{fig7}
\end{center}
\end{figure}

\noindent
We consider in terms of quantitative analysis data for the maximum of the 
average specific heat [Fig. \ref{fig7}a] and the maximum of the average 
susceptibility [Fig. \ref{fig7}b], as functions of chain lengths $N$. 
Using the law $\overline{c_\epsilon(T,N)}^{max}\propto  N^{\alpha_r/\nu_r}$,
which is a particular case of (\ref{maxlaw}), we obtain from the
analysis of these data $\alpha_r>0$ and correspondingly $\nu_r<2$ 
when considering only the smallest chain lengths. In detail, the exponent 
would be still compatible with the value $\nu_r=\nu_p=1$ characterizing a 
first order transition, for $N \le 1000$. For both 
$\overline{c_\epsilon(T,N)}^{max}$ and $\overline{\chi_\epsilon(T,N)}^{max}$ 
the asymptotic saturation becomes obvious only for sizes larger than 
$N \sim 5000$. Following \cite{WiDo}, we consider a fit of the data to the 
form $g_1-g_2N^e$, with $g_1,g_2>0$ and where the exponent is 
$e_c=\alpha_r/\nu_r$ for the specific heat and $e_\chi=\gamma_r/\nu_r$ for the 
susceptibility. Using the whole data sets in the fits, negative exponents, 
compatible with zero within the errors, are obtained in both cases. On the 
other hand, with corrections to scaling, strictly negative exponents are 
obtained. Letting aside the two smallest sizes ($N=100$ and $N=200$),
the best fit in the case of $\overline{c_\epsilon(T,N)}^{max}$
corresponds to $e_c=\alpha_r/\nu_r=-0.3 \pm 0.1$, and we get a compatible 
value for $e_\chi=\gamma_r/\nu_r$ from $\overline{\chi(T,N)}^{max}$. 
This result confirms that $\alpha_r=\gamma_r$ and, from the hyperscaling 
relation, it implies $\nu_r \sim 3$. Intriguingly, the correlation
length exponent value is close to the one obtained for the 
disordered PS model with $c_p=1.75$ considered in \cite{MoGa}. We notice that 
limiting the analysis to $N>500$ we obtain a still larger $\nu_r$, but 
the statistics in our study do not allow more accurate evaluations. 
Nevertheless, the important point is that, when looking at average quantities, 
it clearly appears that the transition is at least of second order and,
at the same time, the crossover between a {\em pure system like} behavior
for small sizes and the (apparent) asymptotic one is quantitatively
confirmed.

\begin{figure}[hptb]
\begin{center}
\leavevmode
\includegraphics*[angle=270,width=11cm]{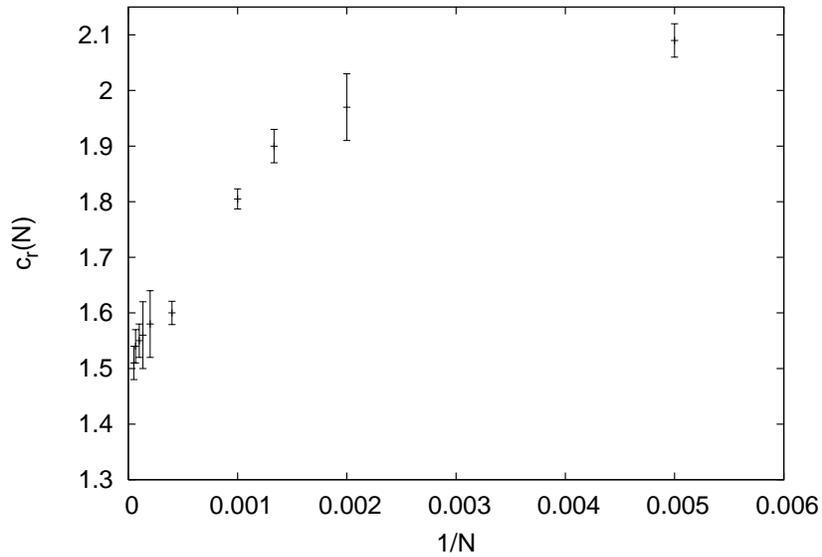}
\caption{Evaluations of the {\em finite size} critical exponent $c_r(N)$, 
plotted as function of $1/N$. The results are from the fits of the probability 
distribution of the loop length to the expected law (\ref{pstartc}) 
$\overline{P^*_\epsilon(l,T_c(N),N)} \propto l^{c_p-c_r}$ (with $c_p=2.15$ and 
$T_c(N)$ taken as the temperature for which the average specific heat is 
maximum). For each chain length, data are fitted in the $l$-range (with $l>2$) 
in which $\overline{P^*(l,T_c(N),N)}$ is increasing with $l$. Here the errors 
are only indicative, as they depend strongly on the number of points in the 
range of the fit.}
\label{fig8}
\end{center}
\end{figure}
\nopagebreak
\begin{figure}[hptb]
\begin{center}
\leavevmode
\includegraphics*[angle=270,width=10cm]{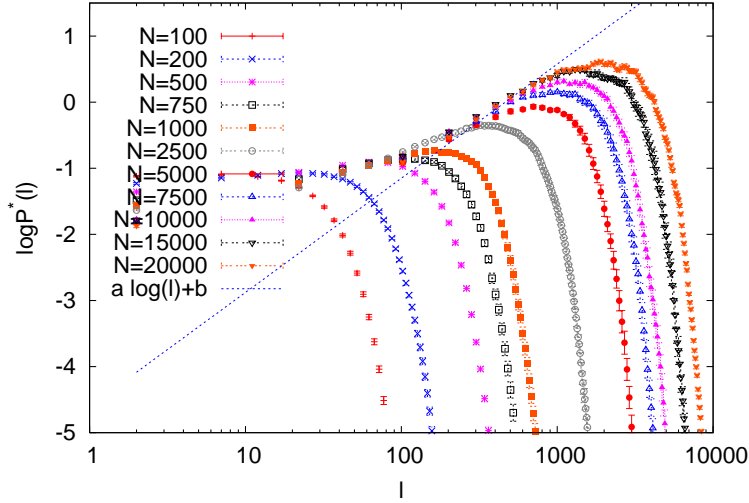}
\caption{For the different chain lengths considered, data for 
$\overline{\log P^*_\epsilon(l,T_c(N),N)}$, at temperatures $T_c(N)$ for which 
average specific heats reach their maxima. The expected asymptotic behavior 
$(2.15-c_r) \log l +b$ with $c_r = 1.4$ (i.e. $c_r \sim 1+1/\nu_r$ with 
$\nu_r$ 
obtained from the fit of the maximum of the specific heat) is also plotted.}
\label{fig9}
\end{center}
\end{figure}

For further validation of this result, and for a better understanding
of finite size corrections to scaling, we consider the fits of 
$\overline{P^*_\epsilon(l,T_c(N),N)}$ to the expected behavior 
$\propto l^{c_p-c_r}$ (\ref{pstartc}). We get correspondingly the
 {\em size-dependent} estimations of the critical exponent $c_r(N)$,
which are presented in [Fig. \ref{fig8}]. Here the (finite size) 
critical temperatures $T_c(N)$ are taken as the temperatures for which the 
average specific heat reaches its maximum and we disregard the possible 
presence of a finite correlation length. We fit data by considering only the 
$l$-range (with $l>2$) in which $\overline{P^*_\epsilon(l,T_c(N),N)}$ is an 
increasing function of $l$ (see [Fig. \ref{fig6}]). The obtained values of 
$c_r(N)$, and therefore of $1/\nu_r(N)=\min \{1,c_r(N)-1\}$, for different 
chain lengths are definitely not compatible within the (though indicative) 
errors and one observes a clear trend towards decreasing $c_r(N)$ values for 
larger chain lengths. It is in particular interesting to notice that, 
for $N=200$, we still have $c_r(N) \simeq 2.1$ and correspondingly 
$\nu_r(N)=1$, whereas for $N\sim 1000$, we obtain 
$c_r(N) \sim 1.8 \div 1.9$, again in perfect agreement with the results 
of \cite{Co} concerning the $3d$ DSAW-DNA on-lattice model. On the other hand, 
with the study of larger chain lengths it becomes clear that the transition is 
at least of second order with $\nu_r \ge 2$. In fact, for the largest size 
considered $N=20000$, the exponent obtained is $c_r(N)\simeq 1.5$. This 
implies that the correct evaluation of the (asymptotic) critical 
exponents, apart from finite size corrections, as
$c_r =\lim_{N \rightarrow \infty} c_r(N)$ leads to $c_r \le 1.5$. 
This value can be compatible with $c_r=1+1/\nu_r \sim 1.35$ 
obtained from the maximum of the specific heat.

It is nevertheless important to stress that the above results 
concern average quantities. The thermodynamic limit behavior of the typical 
sample could be therefore blurred by the fluctuations of the 
sequence-dependent pseudo-critical temperature, if they are governed by an 
exponent $\nu_{r,2}>\nu_{r,1}$ as suggested in \cite{GaMo,MoGa,Mopr,MoGapr}. 
It is possible that we are  only looking at the {\em average} 
correlation length $\xi_2(T,N)$, and in particular the value $c_r=1.5$ 
would be in agreement with the result $\nu_{r,2}=2$ in \cite{MoGa}.

Accordingly, we consider also data on 
$\overline{\log P^*_\epsilon(l,T_c(N),N)}$, evaluated by averaging over 
disorder after taking the logarithm. To be precise, following \cite{MoGa}, the 
quantity expected to be described by the typical correlation
length $\xi_1$ is 
$\sum_i \log \{[Z^f_\epsilon(i,N)Z^b_\epsilon(i+l,N)]/Z^*_{\epsilon,N} \}$. 
Here we are instead considering a kind of mixed average, but the 
behavior of $\overline{\log P^*_\epsilon(l,T_c(N),N)}$ should anyway display 
differences with that of $\log \overline{P^*(l,T,N)}$ in the presence 
of two distinguishable correlation lengths. On the contrary, the comparison 
between [Fig. \ref{fig9}] and [Fig. \ref{fig6}] shows that, at least for 
$T=T_c(N)$, $\log{\overline{P_\epsilon}}$ and $\overline{\log{P_\epsilon}}$ 
behave similarly. In particular, $\overline{\log{P^*_\epsilon}}$ is also an 
increasing function of $l$ for the largest considered sizes, and on 
quantitative grounds the evaluated $c_r(N)$ values are essentially compatible 
within errors. In order to emphasize this point, the expected asymptotic 
behavior $\overline{\log{P^*_\epsilon}} \sim (2.15-c_r) \log l+b$ with 
$c_r = 1+1/\nu_r \sim 1.35$ (obtained from the average specific heat) is also 
plotted in [Fig. \ref{fig9}]. More quantitative analysis of these results are 
left for a forthcoming work \cite{CoYe2}.

\section{Conclusions}
\noindent
We studied numerically a disordered PS model for DNA denaturation with 
$c_p=2.15$, which displays a first order transition in the homogeneous case, 
by solving recursively the equations for the canonical partition 
function with the SIMEX scheme. The model is made as similar as possible
to the $3d$ DSAW-DNA previously studied by Monte Carlo like simulations
\cite{Co} and it is expected that the results of the study 
could also be relevant to different disordered PS models with $c_p>2$. 

We introduced the parameter $x=R/[\log \mu (R-1)]$, where 
$R=\epsilon_{GC}/\epsilon_{AT}$ is the ratio of the Guanine-Cytosine to the 
Adenine-Thymine coupling energies and $\mu$ is the connectivity constant of 
the corresponding on-lattice model. We showed that this parameter, at fixed 
$c_p$ value and GC composition (taken to be 1/2), appears to play the role of 
(the logarithm of) an {\em intrinsic} length scale, and to describe, in a 
first approximation, the finite size behavior. In particular, for a given 
$N$-value, the manifestation of the effect of disorder appears to be the most 
evident for the smallest $x$ values, in agreement with a qualitative 
explanation based on the possible occurrence of large enough {\em rare 
regions}. It is interesting to notice that, within this picture, the 
system size necessary for observing the asymptotic behavior and making
evident the effect of disorder diverges exponentially with $x$.

We studied in detail the value $x=1.3$, obtained with $R=2$ 
and $\log \mu=1.54$ (as in \cite{Co}), for sequence sizes up to $N=20000$ 
(larger than the sizes accessible to Monte Carlo like simulations by a factor 
20). We found that the model exhibits strong corrections to scaling,
displaying a crossing between a still nearly {\em pure system like} behavior 
for small chain lengths $N \lesssim 1000$ and the observed (apparently 
asymptotic) large one for $N \gtrsim 5000$. In particular, the maximum of the 
average specific heat, which behaves as the susceptibility, increases with $N$ 
for small chain lengths. Considering the whole size range, it appears instead 
to be clearly saturating. This result shows that, at least from the point of 
view of average quantities, the thermodynamic limit is described by a random 
fixed point with $\alpha_r \le 0$ and correspondingly $\nu_r \ge 2/d=2$. By 
fitting data with a scaling law of the form $g_1-g_2N^{\alpha_r/\nu_r}$, and 
by taking into account corrections to scaling, we find in particular 
$\alpha_r/\nu_r = -0.3 \pm 0.1$, which by using the hyperscaling relation 
gives $\nu_r \sim 3$.

The average loop length probability distribution at the critical temperature 
appears still described by a power law at least on an interval 
$1 \ll l \ll N$ of the range. Upon fitting data according to  
$\overline{P_\epsilon(l,T_c(N),N)}\propto 1/l^{c_r}$ one finds $N$-dependent 
values for the exponent which is compatible with $c_p$ for the smallest sizes
whereas when looking at the whole $N$ range it appear to converge towards an 
asymptotic limit $c_r \le 1.5$ (possibly compatible with 
$c_r=1+1/\nu_r \sim 1.35$). Moreover, $\overline{\log P(l,T_c(N),N)}$ exhibits 
also a similar behavior, suggesting that there is no difference between 
{\em typical} and {\em average} correlation lengths. 

Our best-fit estimation, $\nu_r \sim 3$, is close to the estimation in 
\cite{MoGa} for the case $c_p=1.75$. This observation would
support the hypothesis that disorder is relevant as soon as
$c_p>3/2$ and that the various disordered PS models considered
could be described by the same random fixed point corresponding to a 
transition which is at least of second order (and probably smoother),
in agreement with recent analytical findings \cite{GiTo1,GiTo2,To}. Our 
statistics do not allow nevertheless to completely rule out the possibility 
that $\nu_r=2$, particularly from $\overline{P_\epsilon}$ data, and in any 
event an analysis in terms of pseudo-critical temperatures 
\cite{WiDo,GaMo,MoGa,MoGa2,Mopr,MoGapr} is in order for clarifying the 
situation. We leave this development to a forthcoming work \cite{CoYe2}.

It is also interesting to notice that there are very recent theoretical
studies \cite{BaKaMu,AmBaLoMe,NoPeAmHaMe} on the loop dynamics in 
PS models, which in particular relates the equilibrium loop length 
distribution probability to the correlation function, therefore suggesting 
a new intriguing method for measuring experimentally the $c$ value.
From this point of view, the expected behavior of $P(l)$ in presence
of disorder and the possibility of observing differences between the average 
and the typical sequence cases seems to us important questions to 
be clarified.

In conclusion, our results provide numerical evidence for strong finite size 
corrections to the asymptotic behavior of the disordered PS model considered. 
The data show moreover that disorder is relevant, at least from the analysis 
of average quantities. The findings here appear also to confirm that
the evaluation $\nu_r \simeq 1.2$ in previous numerical study concerning 
on-lattice $3d$ DSAW-DNA model \cite{Co} is to be considered as lower bound 
for the correct (average) correlation length exponent. The observed behavior 
is in agreement with a proposed qualitative picture for finite size effects, 
which could also explain the difference with the results of previous studies 
on a different disordered PS model with the same $c_p$ \cite{GaMo,MoGa}. A 
preliminary presentation for part of the findings and hypotheses here can be 
found in \cite{CoYepr}.

\section*{Acknowledgments} 
\noindent
B.C. would like to acknowledge an enlightening discussion that she had 
with David Mukamel some time ago. We are moreover grateful to Thomas 
Garel, Cecile Monthus, Andrea Pagnani and Giorgio Parisi for comments.

\section*{Appendix}
\noindent
We use in the numerical computations the SIMEX implementation of the FF scheme
\cite{FiFr,Yeetal,Ye1}, which relies on the numerical approximation of
the powers $1/l^{c_p}$ by sums of exponentials:
\begin{equation}
\frac{1}{(2l)^{c_p}} \simeq \sum_{k=1}^{N_{S}} a_k \exp(-2 l b_k).
\end{equation}
In the present study we consider $c_p=2.15$ and the values for the 
coefficients $a_k$ and $b_k$, with $N_{S}=15$, provided in \cite{GaMo}.
The computation of the recursive equations for the forward and backward 
partition functions were implemented with the introduction of 
{\em free energy like quantities}, in order to handle logarithms of 
$Z^f_\epsilon$ and $Z^b_\epsilon$ (following \cite{GaOr}). 

In detail, the equation (\ref{zf}) for the forward partition function
becomes:
\begin{equation}
Z^f_\epsilon(\rho+1)=\exp({\beta \epsilon_{\rho+1}-\log\mu}) \left \{
Z^f_\epsilon(\rho)+ \sum_{k=1}^{N_{FF}} a_k \sum_{\rho'=1}^{\rho-1} 
{Z^f_\epsilon(\rho')} \exp[-2(\rho-\rho'+1)b_k] \right \}.
\end{equation}
By defining
\begin{equation}
Q_k(\rho)=\sum_{\rho'=1}^\rho Z^f_\epsilon(\rho') \exp(2 \rho' b_k) = 
\exp[2 \rho b_k+\mu_k(\rho)],
\end{equation}
one obtains
\begin{equation}
Z^f_\epsilon(\rho)=\exp[\mu_k(\rho)]-\exp[\mu_k(\rho-1)-2 b_k]
\end{equation}
and a recursion relation for $\mu_k(\rho)$:
\begin{eqnarray}
\mu_k(\rho+1)&=&\mu_k(\rho)+\log(A_f+B_f+C_f) \\
A_f&=&\exp(-2 b_k) \nonumber \\
B_f&=&\exp(\beta\epsilon_{\rho+1}-\log \mu) \left \{
1-\exp[-2b_k+\mu_k(\rho-1)-\mu_k(\rho)] \right \} \nonumber \\
C_f&=&\exp(\beta\epsilon_{\rho+1}-\log \mu) \left \{ \sum_{j=1}^{N_{FF}}
a_j \exp[-4b_j+\mu_j(\rho-1)-\mu_k(\rho)] \right \} \nonumber 
\end{eqnarray}
Analogously, one writes the equation for the backward partition function:
\begin{equation}
Z^b_\epsilon(\rho-1)=\exp({\beta \epsilon_{\rho-1}-\log\mu}) \left \{
Z^b_\epsilon(\rho)+ \sum_{k=1}^{N_{FF}} a_k \sum_{\rho'=\rho+1}^{N} 
{Z^b_\epsilon(\rho')} \exp[-2(\rho'-\rho+1)b_k] +1\right \},
\end{equation}
and defines
\begin{equation}
R_k(\rho)=\sum_{\rho'=\rho}^N Z^b_\epsilon(\rho') \exp(-2 \rho' b_k) = 
\exp[-2 \rho b_k+\nu_k(\rho)],
\end{equation}
obtaining
\begin{eqnarray}
\nu_k(\rho-1)&=&\nu_k(\rho)+\log(A_b+B_b+C_b+D_b) \\
A_b&=&\exp(-2 b_k) \nonumber \\
B_b&=&\exp(\beta\epsilon_{\rho-1}-\log \mu) \left \{
1-\exp[-2b_k+\nu_k(\rho+1)-\nu_k(\rho)] \right \} \nonumber \\
C_b&=&\exp(\beta\epsilon_{\rho-1}-\log \mu) \left \{ \sum_{j=1}^{N_{FF}}
a_j \exp[-4b_j+\nu_j(\rho+1)-\nu_k(\rho)] \right \} \nonumber \\
D_b&=&\exp(\beta\epsilon_{\rho-1}-\log \mu)\exp[-\nu_k(\rho)]. \nonumber 
\end{eqnarray}
We used the boundary conditions (with the implicit assumption
$Z^f_\epsilon(0)=Z^b_\epsilon(N+1)=0$):
\begin{eqnarray}
Z^f_\epsilon(1)&=&\exp(\beta \epsilon_1-\log \mu)\nonumber \\ 
Z^f_\epsilon(2) &=& \exp(\beta \epsilon_1-\log \mu)
\exp(\beta \epsilon_2-\log \mu)\nonumber\\
Z^b_\epsilon(N-1)&=&\exp(\beta \epsilon_{N-1}-\log \mu)
[\exp(\beta \epsilon_{N}-\log \mu)+1] \nonumber \\
Z^b_\epsilon(N) &=&\exp(\beta \epsilon_N-\log \mu), 
\end{eqnarray}
and correspondingly:
\begin{eqnarray}
\mu_k(1)&=&\beta \epsilon_1-\log \mu \nonumber \\
\mu_k(2)&=&\beta \epsilon_1-\log \mu
+\log[\exp(-2b_k)+\exp(\beta \epsilon_2-\log \mu)] \nonumber \\
\nu_k(N-1)&=&\beta \epsilon_N-\log \mu
+\log \left \{ \exp(-2b_k)+\exp(\beta \epsilon_{N-1}-\log \mu)
[1+\exp(-\beta \epsilon_{N}+\log \mu)] \right \}\nonumber \\
\nu_k(N)&=&\beta \epsilon_N-\log \mu. 
\end{eqnarray}

\end{document}